\documentclass[apj]{emulateapj}
\usepackage{float}
\usepackage{amsmath}
\usepackage{multirow}
\bibpunct{(}{)}{;}{a}{}{,}

\begin{document}
% Title Page
\title{Quasi-periodic pulsations in solar and stellar flares: re-evaluating their nature in the context of power-law flare Fourier spectra}

\author{A. R. Inglis\altaffilmark{1}, J. Ireland\altaffilmark{2}}
\affil{Solar Physics Laboratory, Heliophysics Science Division, NASA Goddard Space Flight Center, Greenbelt, MD, 20771, USA}

\author{M. Dominique}
\affil{ Solar-Terrestrial Center of Excellence, Royal Observatory of Belgium, Avenue Circulaire 3, B-1180, Brussels, Belgium}
\altaffiltext{1}{Physics Department, The Catholic University of America, Washington, DC, 20664, USA}
\altaffiltext{2}{ADNET Systems Inc.}

\begin{abstract}
The nature of quasi-periodic pulsations in solar and stellar flares remains debated. Recent work has shown that power-law-like Fourier power spectra, also referred to as `red' noise processes, are an intrinsic property of solar and stellar flare signals, a property that many previous studies of this phenomenon have not accounted for. Hence a re-evaluation of the existing interpretations and assumptions regarding QPP is needed. Here we adopt a Bayesian method for investigating this phenomenon, fully considering the Fourier power law properties of flare signals. Using data from the PROBA2/LYRA, Fermi/GBM, Nobeyama Radioheliograph and Yohkoh/HXT instruments, we study a selection of flares from the literature identified as QPP events. Additionally we examine optical data from a recent stellar flare that appears to exhibit oscillatory properties. We find that, for all but one event tested, an explicit oscillation is not required in order to explain the observations. Instead, the flare signals are adequately described as a manifestation of a power law in the Fourier power spectrum, rather than a direct signature of oscillating components or structures. However, for the flare of 1998 May 8, strong evidence for the existence of an explicit oscillation with $P \approx$ 14-16 s is found in the 17 GHz radio data and the 13-23 keV Yohkoh HXT data. We conclude that, most likely, many previously analysed events in the literature may be similarly described in terms of power laws in the flare Fourier power spectrum, without the need to invoke a narrowband, oscillatory component. As a result the prevalence of oscillatory signatures in solar and stellar flares may be less than previously believed. The physical mechanism behind the appearance of the observed power laws is discussed.
\end{abstract}

\maketitle

\section{Introduction}
\label{intro}

A common feature of solar flare emission is the appearance of quasi-periodic pulsations (QPP). Alternately known - particularly in stellar and astrophysical contexts - as quasi-periodic oscillations (QPO), these phenomena have been observed in a wide range of wavelengths over several decades \citep[see][for a recent review]{2009SSRv..149..119N}. Although not precisely defined in the literature, the term QPP is most often used to describe variations in the flux from a flare or other astrophysical object as a function of time, which appear to include periodic components with characteristic timescales ranging from one second up to several minutes. They are typically observed during the impulsive phase of solar flares and have been observed over a wide range of wavelengths, from radio waves and microwaves to hard X-rays and gamma-rays. Similar signatures have also been observed from stellar flares \citep[e.g.][]{2006A&A...456..323M, 2010ApJ...714L..98K}. Since QPP are directly linked
to the properties of the flare reconnection region and flare acceleration sites, a full description of QPP remains crucial for our understanding of solar flares.

There are two main theories that are currently being pursued as possible mechanisms for generating QPP \citep[see][]{2009SSRv..149..119N}, both of which assume the presence of a periodic driver. These are that the observed flux variations are driven either by (1) magnetohydrodynamic (MHD) wave behavior in the corona and in flare sites, or are instead (2) a result of a regime of periodic or ‘bursty’ reconnection. As a consequence, recent studies have focused on the assumed periodic nature of QPP; such studies are usually motivated by the concept of searching for a periodic signal obscured by random noise and long-term trends in the flare signal. 

However, it has recently become clear \citep[e.g.][]{2011A&A...533A..61G, 2010MNRAS.402..307V} that flare time series are often dominated by a power law in the Fourier domain, rather than random white noise. In an astrophysical context this is often referred to as `red noise'.
Objects that are known to exhibit time series with power-law like behaviour in the Fourier power spectrum include XMM-Newton observations of Seyfert galaxies \citep{2005A&A...431..391V, 2010MNRAS.402..307V}, gamma-ray bursts \citep{2010AJ....140..224C}, active galactic nuclei \citep{2006Natur.444..730M}, and magnetars \citep{2013ApJ...768...87H}. The Fourier power spectra of all of these objects are well-described using power law models with a negative slope, i.e., $P(f) \approx f^{-\alpha}$ for $\alpha \geq 0$ where $f$ is frequency. Strictly speaking, `red noise' refers to a specific slope of the Fourier power law spectrum, $\alpha$ = 2, however the term is often more loosely used to refer to any power-law dominated Fourier power spectrum where $\alpha \geq 0$. The term `noise' in this context is also something of a misnomer; the observed emission is not noise in the conventional sense of detector noise or measurement error. Rather, the power-law shape of the Fourier power spectrum is an intrinsic property of the physical system and must therefore be taken into account when examining for other effects such as oscillations. 

That solar flares exhibit Fourier power-law-like properties has in fact been known for some time \citep[e.g.][]{1997A&A...324..750R, 1998ApJ...505..941A, 1998A&AS..127..309S}. Such properties are clearly present in the Fourier power spectra of RHESSI X-ray flare observations. However, until recently the implications for QPP have not been considered. \citet{2011A&A...533A..61G} were the first to demonstrate the importance of considering this property of flaring emission in QPP studies by analyzing Fermi/GBM and RHESSI solar flare data from the same events in two different ways. They demonstrated that the consequence of applying a white-noise assumption to data dominated by a power law in the Fourier power spectrum is a drastic overestimation of the significance of peaks detected in the Fourier domain. Additionally, the nature of the Fourier power spectrum in such signals means that the empirical subtraction of `background components' of the signal may lead to misleading results and should be avoided; the entire Fourier spectrum of the signal should always be considered \citep{2010MNRAS.402..307V}.

Hence the results of a number of previous studies, and the general prevalence of flare signals with oscillatory content, are both called into question. However, as \citet{2011A&A...533A..61G} also point out, their study does not mean that observed variations in flare signals are not real, and we reiterate that here. Rather, it means that the physical mechanisms causing the observed flux variations need not be exclusively due to an oscillatory mechanism. 

In this paper, we address this problem and the question of the true prevalence of statistically significant narrowband oscillations in solar flares. We achieve this by adopting an approach based on \citet{2010MNRAS.402..307V}, which takes into account the true statistical properties of flare time series. In this paper, we introduce the method and demonstrate its application to selected solar and stellar events where the presence of QPP has been suggested.

\section{Instruments and data selection}

For this study we concentrate on a joint dataset consisting of X-ray data from the Gamma-ray Burst Monitor (GBM) \citep{2009ApJ...702..791M} on board Fermi, and soft X-ray and EUV data from the LYRA instrument \citep{2013SoPh..286...21D} on board PROBA2. Fermi was launched in 2008 and GBM, which consists of 12 NaI and 2 BGO detectors, has provided high-cadence solar observations in the 5 keV - 40 MeV energy range since then. Up to six of the NaI detectors and one BGO detector may be sunward facing at any given time.  

PROBA2 was launched in 2009 with a science payload including the Large Yield Radiometer (LYRA), which  observes the Sun in 4 channels with a nominal time resolution of 0.05 s, although the instrument is capable of 0.01 s resolution. Of these channels, two continue to take science quality data, namely the  Aluminium (Al) filter (17-80 nm + SXR) and the Zirconium (Zr) filter (6-20 nm + SXR). The availability of these simultaneous datasets provides us with the necessary multi-wavelength information for analysing QPP effectively in different regimes, as well as the time cadence required for the detection of pulsations in the 1 - 300 s range.

Two recent solar flares considered to be QPP events are the GOES-class X2.2 flare of 2011 February 15 \citep[e.g.][]{2012ApJ...749L..16D}, and the GOES-class M2.5 flare of 2011 June 7 \citep[e.g.][]{2013ApJ...777...30I}. Both of these events were fully observed by the LYRA and GBM instruments, making them ideal candidates for a multi-wavelength investigation of their power spectral properties.

For each of these events we apply the analysis method described in Section \ref{method} to a selection of GBM data channels. For this analysis we utilize the CTIME data product, which provides X-ray count information at a 0.256s temporal cadence in nominal observing mode, and $\approx$ 0.064s cadence during flare times. GBM CTIME data is available in 8 energy channels, covering the range 4 keV - 2 MeV. For the analysis presented in this paper we select the following intervals: Channel 1 ($\approx $ 12 - 27 keV), Channel 2 ($\approx$ 27 - 50 keV) and Channel 3 ($\approx$ 50 - 100 keV). The lowest energy emission observed in Channel 0 (4-12 keV) is not used. These intervals are suited to capturing both the soft and hard X-ray emission from flares. To ensure a consistent observational cadence and to improve signal-to-noise, all the analyzed GBM data are re-binned into 1s intervals prior to analysis. Similarly, for LYRA the data are accumulated into 1s intervals throughout this paper. Subsequently, identical analysis is applied to data obtained from the GBM channels to capture hard X-ray behaviour, and to the Al and Zr filters, capturing EUV and soft X-ray emission.  Hence the result will be a thorough understanding of the Fourier spectral properties of these flares across EUV, soft X-ray and hard X-ray wavelengths.

Additionally, we return to the flare of 1998 May 8th, previously investigated in the microwave and X-ray regimes by \citet{2004AstL...30..480S} and \citet{2008A&A...487.1147I}, perhaps one of the most pronounced examples of QPP in the prior literature. During this flare, pronounced pulsations were observed in the 17 GHz data from the Nobeyama Radioheliograph (NoRH) \citep{1994IEEEP..82..705N}, as well as in X-ray flux in the 13-93 keV range as observed by the Yohkoh satellite \citep{1991SoPh..136....1O}. In \citet{2004AstL...30..480S, 2008A&A...487.1147I} the observed pulsations were analysed using a white-noise assumption, resulting in an observed 16s period which was interpreted as a signature of either a magnetohydrodynamic ballooning mode or a magnetoacoustic sausage mode. However, it is timely to revisit this event and re-analyse the Nobeyama Radioheliograph (NoRH) radio data and the Yohkoh X-ray data in the context of power law Fourier spectra.

Finally, for a stellar flare perspective we analyse near-UV U-band ($\sim$ 3200$\AA$ - 3900$\AA$) data from the New Mexico State University (NMSU) 1m telescope \citep{2010AdAst2010E..46H} for the `megaflare' of 2009 January 16, previously studied by \citet{2010ApJ...714L..98K, 2013ApJS..207...15K, 2013ApJ...773..156A}. This flare exhibited interesting temporal variations which were recently investigated by Anfinogentov et al. (2013), who suggested that the observed emission was the manifestation of a magnetoacoustic wave. However, as with the solar flares above, the nature of the Fourier power spectrum was not considered in the analysis of this event.

\section{Data preparation}

\subsection{LYRA Large-angle rotations (LARs)}
\label{lars}

In order to apply our analysis method to solar flare data, the GBM and LYRA data require some preparation. In particular,
a design feature of PROBA2 is the incidence of large-angle rotations (LARs), in which the spacecraft is rotated approximately 90 degrees  in order to accommodate the on-board star tracker. These manoeuvres are performed 4 times per orbit, or approximately every 25 minutes. In some cases, a consequence of this is the appearance of temporal artifacts in the LYRA data during LARs, due to the movement of the Sun within the instrument's field of view. 

\begin{figure}
\begin{center}
\includegraphics[width=8cm]{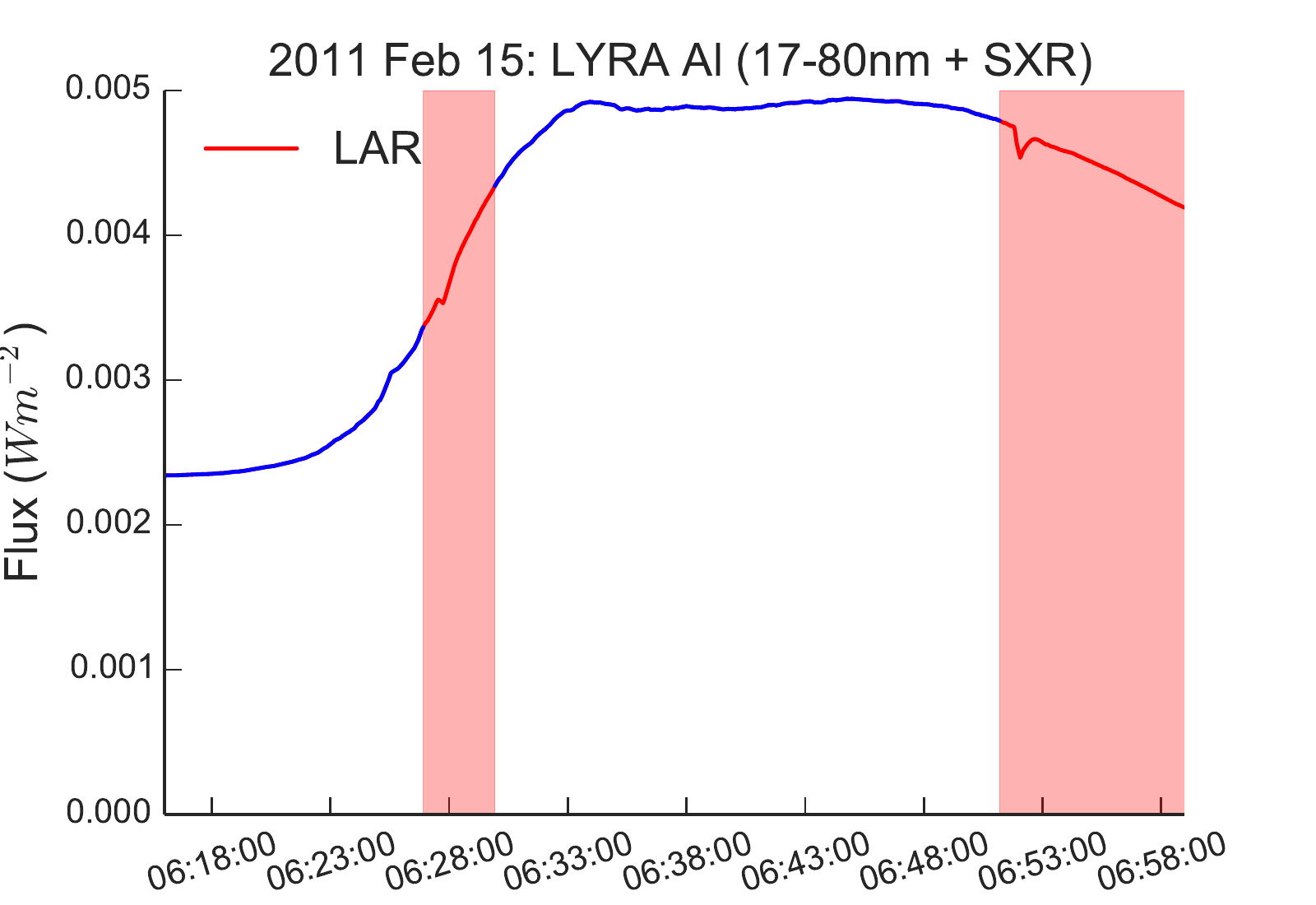}
\caption{Example of LYRA large angle rotations (LARs) - the areas in red are marked as affected by the manoeuvre by the LYRA timeline annotation file.}
\label{data_prep}
\end{center}
\end{figure}

A list of LARs and other spacecraft events (for example, passes through the South Atlantic Anomaly) is maintained by the PROBA2 instrument team, known as the \textit{LYRA timeline annotation file (LYTAF)}. In this work we remove the effects of LARs in the simplest way, by removing the affected times from the data entirely. This is achieved using routines within the SunPy \citep{mumford-proc-scipy-2013} data analysis package, which query the LYTAF, and, for a given data set, automatically separate a LYRA time series into components, excluding the affected portions of the data (see Figure \ref{data_prep}).

%\subsection{GBM data dropouts}
%GBM does not experience noticeable rotational artifacts like LYRA (or RHESSI), however the NaI detectors experience occasional dropouts which lead to discontinuities in the flare signal. Such artifacts would have a strong effect on the Fourier spectrum of the data. Hence, we filter these dropouts via an interpolation method, which replaces the affected data points with an interpolated version of the signal (see Figure \ref{data_prep}).  The signal affected by these dropouts usually spans no more than 3-4 data points at any one time.

\subsection{Data normalisation and apodizing}

An additional step in data preparation for all data sources is to normalise and apodize the input signal $F$. This normalisation is done according to,

\begin{equation}
F_{norm} = \frac{F - \bar{F}}{\bar{F}}
\label{norm_eqn}
\end{equation}

where $\bar{F}$ is the mean of the signal. Finally, the normalised signal is prepared by multiplying it by a window function. This step is crucial, otherwise the Fourier PSD of the data is dominated by the discontinuity caused by the mismatch between the beginning and end of the signal. For this purpose we choose the well-stablished Hann window \citep{Blackman:1959:MPS}. For verification, other window functions (e.g. Blackman-Harris, Hamming) were tested and found to produce very similar results; for brevity these tests are not shown in this paper.

In all of the results shown in Section \ref{results}, the analysis method has been applied to the prepared, apodized data as described here.

\section{The method}
\label{method}

A powerful and flexible methodology designed for testing the presence of oscillations in the presence of frequency-dependent noise is described in \citet{2010MNRAS.402..307V}, where it is applied to determine the presence of quasi-periodic oscillations in XMM-Newton observations of highly variable Seyfert 1 galaxies, objects that also exhibit frequency-dependent Fourier power spectra. A key advantage of this approach is that we avoid techniques subtracting empirically defined background trends, which inevitably modify the frequency content of the analyzed data \citep{2010AJ....140..224C, 2014A&A...563A...8A}. Additionally, one of the main motivations for adopting this Bayesian methodology is the enablement of posterior predictive checking, as described in Section \ref{method_step3}. Finally, Bayesian methodology also makes it straightforward to incoroporate any prior information we may have into the procedure, and to easily test different models of the observed power spectrum. 

The \citet{2010MNRAS.402..307V} approach is a multi-stage process. In the first stage (see Section \ref{method_step1}), Bayes' theorem is used to fit parameterized models $S$ (e.g. a power law) to the Fourier spectral power density. This is achieved via a Markov-Chain Monte-Carlo (MCMC) procedure, which explores the parameter space of the chosen model to find the posterior probability distribution. From this the model that best fits the data may be extracted.

The next stage is to perform a model comparison, to establish which of the tested models $S$ is most favoured. Numerous methods are available to perform this model selection procedure. Having selected the most favoured model, appropriate test statistics are calculated (Section \ref{method_step2}). Then, (see Section \ref{method_step3}) posterior predictive checking is performed to establish the distribution of the relevant test statistics. This allows us to establish whether the measured test statistics were extreme, and therefore whether the chosen model captures the statistics of the data. Hence, by choosing relevant models, this procedure can test whether a model that includes an oscillatory component is needed in order to explain the observed power spectra of solar and stellar flares.

In the following sections we describe in brief the methodology used in this paper, which is similar to that used in \citet{2010MNRAS.402..307V}. We begin with Bayes’ Theorem.

\subsection{Determining the model parameters}
\label{method_step1}

Given that $p(a|b)$ denotes the probability of $a$ given $b$ then Bayes’ Theorem states that

\begin{equation}
p(H|D,I) = \frac{p(H|I) p(D|H,I)}{p(D|I)}
\label{bayes_theorem}
\end{equation}

where $H$ represents the hypothesis to be tested (in our context a model of the power spectral density), $D$ is the data (in our case the measured power spectral density), and $I$ represents the prior information available. The quantity $p(D|H, I)$ is called the likelihood, and is the probability of the data $D$ given the hypothesis $H$. The quantity $p(H|I)$ is called the prior information, and relates what we know about the hypothesis from our previous information before we take any data. The denominator $p(D|I)$ represents the prior probability of the data, and is a constant. The left-hand side - $p(H|D,I)$ - is the posterior probability of the hypothesis after we have acquired some data, and expresses all of what we know about the hypothesis $H$ \citep[see][]{2010blda.book.....G}.

\citet{2010MNRAS.402..307V} uses Bayes' Theorem to compare hypotheses - or models - of red-noise spectra to the observed power spectral density (PSD) of Seyfert 1 galaxies. The same treatment may be applied to solar flares. A model spectral power density $S_j=S(f_j,\Theta)$ is chosen, where $f_j$ is the Fourier frequency, and $\Theta=(\theta_1,....\theta_m)$ represents the $m$ parameters of the model (for example, the parameters of a simple power law model consists of a power law index and a normalization constant). The periodogram of any observed stochastic time series of length $N$ has Fourier power $D^{obs}$ = ($D^{obs}_{1},...,D^{obs}_{N/2}$) at Fourier frequency $f_j = j/N\Delta T$ (with $j=1, . . . , N/2$) and is exponentially distributed (Press et al 1992; Chatfield 2003) about the true spectral density $S_j = S(f_j)$. This allows us to write the likelihood as

\begin{equation}
p(D^{obs}|\Theta,I) = \prod^{N/2}_{j=1} \frac{1}{S_j} \exp \left( - \frac{D_j^{obs}}{S_j} \right)
\end{equation}

where we have replaced the hypothesis $H$ with its equivalent representation $\Theta$. 

The posterior $p(H | D, I)$ can be calculated as a function of its parameters $\Theta$ using a MCMC method. In this work, the posterior distributions are calculated using \verb|PyMC|, a Python analysis package which implements such an MCMC procedure. MCMC methods allow for the efficient mapping of Bayesian posterior probability density functions in multi-dimensional parameter space. After some initial period (known as `burn-in'), the Markov chain returns samples of the parameters $\Theta$ directly proportional to their probability density as defined by the Bayesian posterior; that is, the equilibrium distribution of the Markov chain is the same as the posterior probability density function (Gregory 2005). Therefore, volumes of the parameter search space containing probable solutions are sampled more often in preference to volumes containing less probable solutions. This Bayesian/MCMC approach calculates the probability density function (PDF) for $\Theta$. This posterior PDF contains a lot of information: for example, we can find the best fit by finding $\Theta_{mode}$, the value of $\Theta$ that has the maximum posterior probability. The posterior PDF can also be used to find the best-fit and probability distribution of each fit parameter \citep[e.g.][]{2013ApJ...769...89I}.

\subsection{Candidate models}

In this work we focus on two candidate models that may describe the power spectral density of flare time series data. First, we consider a single power law model plus a constant, i.e.

\begin{equation}
S_A = A f^{-\alpha} + C
\end{equation}

This model is physically motivated by the understanding that power-law Fourier power spectra are a common feature of astrophysical objects, as discussed in Section \ref{intro}. The addition of the constant $C$ accounts for the transition between a power law (`red') regime to a white-noise regime or a Poisson regime, as observed by \citet{2010AJ....140..224C} among others.

The second model includes an additional component that accounts for the appearance of an oscillatory component in the data, in addition to any observed power laws. This model may be written,

\begin{equation}
S_B = S_A + B \exp \left( \frac{-(\ln f - \beta)^2}{2\sigma^2} \right)
\end{equation}

where $B = B_0 / \sqrt{2\pi\sigma^2}$, such that the integral over the Gaussian component is equal to 1. Hence model $S_B$ is model $S_A$ with a function added that is equivalent to a Gaussian in log-frequency space. The width of this Gaussian is then given by $\sigma$, its location in log-frequency by $\beta$, and it's amplitude by $B$. This function represents excess power in the signal that may arise due to an oscillation.

\subsection{Selection of priors}

In Bayesian analysis, the choice of prior probabilities affects the final posterior probability distribution (Equation \ref{bayes_theorem}). These priors represent the information we already possess about the model characteristics. In this work, we adopt uniform prior probabilities for the model parameters:

\begin{equation}
\begin{split}
 -10 < \log A < 10, \\
 -6 < \log B_0 < 5, \\
 -20 < \log C < 10, \\
 -1 < \alpha < 6, \\
 -6.0 < \beta < -2.0, \\
 0.05 < \sigma < 0.25, \\
\end{split}
\end{equation}
 
Hence we define the prior probability of each parameter to be uniform within a set range, i.e. any value within that range is equally likely. This intentionally simple choice reflects our lack of precise prior knowledge about the signal and ensures that we do not accidentally exclude the optimal values of the model parameters from the analysis.

\subsection{Model Selection}

In order to determine which of the candidate models is a preferred fit to the data, we use the Bayesian Information Criterion \citep[BIC;][]{schwarz1978, Burnham01112004}. Similarly to log-likelihood measurements, the BIC is minimized in the process of finding the best-fit of the model parameters to the data. The BIC criterion is defined as,

\begin{equation}
BIC = -2 \ln (L) + k \ln (n)
\label{bic_eqn}
\end{equation}

for large $n$, where $n$ is the number of data points, $k$ is the number of parameters in the model, and $L$ is the maximum likelihood. 

%Similarly, AIC is defined as,

%\begin{equation}
%AIC = 2k - 2\ln (L)
%\label{aic_eqn}
%\end{equation}

A feature of the BIC is that, as Equation \ref{bic_eqn} shows, it includes built-in consideration of the number of parameters $k$ in the model. Models with more parameters are penalized in comparison to those with fewer parameters, to compensate for that fact that the more complex model should always fit at least as well as the simple one.

In this paper we will use the BIC to perform the model comparison. By comparing the BIC of two models one may find the extent to which one model is preferred over the other, e.g.:

\begin{equation}
\Delta BIC = BIC_{A} - BIC_{B}
\end{equation}

Since lower values of BIC are preferred, a negative value of $\Delta BIC$ indicates that model $S_A$ is preferred over model $S_B$, whereas a positive value indicates a preference for $S_B$. In general, a value of $\pm 10$ in $\Delta BIC$ is considered highly significant \citep[see e.g.][]{Burnham01112004} . %From Equation \ref{bic_eqn} this may described in terms of probabilities as:

%\begin{equation}
%\frac{p_1}{p_0} = \exp \left(\frac{\Delta BIC}{2} \right)  
%\end{equation}

%For example, if $\Delta AIC$ = -3, then we find that $p_1/p_0 \approx $ 0.22, indicating that $S_B$ is less probable than $S_A$. 
It is important to note however, that the BIC merely measures which of the tested models is more appropriate, or alternatively which model minimizes information loss. However, it does not explicitly test whether either model is a good choice in absolute terms - both models may be poor fits to the observed data. In order to determine this, additional test statistics are required, as described in Section \ref{method_step2} below.
  
\subsection{Test statistics}
\label{method_step2}

In addition to model selection via the BIC, we include other measures of the appropriateness of the chosen models. We consider two test statistics, defined as follows \citep{2010MNRAS.402..307V}:

\begin{equation}
T_{SSE} = \sum_j \left( \frac{D_j - S_j}{S_j} \right)^2
\end{equation}

and,

\begin{equation}
T_R = \max_j \left( 2 D_j / S_j \right)
\end{equation}

The statistic $T_{SSE}$ corresponds to a global goodness-of-fit measure of the model $S$ to the observations $D$, similar to $\chi^2$. $T_R$ measures the maximum deviation of the observations $D$ from the model $S$. Hence $T_R$ is a valuable statistic for finding local anomalies between $I$ and $S$, such as might occur due to a narrowband oscillation.

The distribution of test statistic values, and hence to derive \textit{p}-values corresponding to the values of $T_R$ and $T_{SSE}$, is achieved via posterior predictive checking, described in Section \ref{method_step3}. This allows a determination of whether the measured values of $T_R$ and $T_{SSE}$ are extreme compared to this distribution, which could indicate a poor choice of model. %In the context of solar flares, the detection of a narrowband oscillation would correspond to a flare where the value of $T_{SSE}$ obtained was not extreme, indicating a good overall fit of the model, but where $T_R$ was extreme, indicating that a significant local spike in the Fourier spectrum was present.

\subsection{Posterior predictive checking: estimating test statistic distributions}
\label{method_step3}

Section \ref{method_step1} gives us the posterior distribution of values for the model $S$ parameters. Sampled parameter values from this posterior can be used to generate simulated noisy spectral power density data $D_{rep}$, using the spectral model. This simulated data can be used to calculate the distribution of any test statistic $T(D), D=(D_1,...,D_{N/2})$. The observed value of the test statistic is $T(D_{obs})$, where $D_{obs}$ is the original data. The probability density distribution $p(T)$ for the test statistic is found by repeatedly generating simulated spectral power density data $D_{rep}$ and calculating $T(D_{rep})$. The Bayesian \textit{p}-value $p_B$ is then defined as the tail area probability that the simulated data could give a test statistic at least as extreme as that observed, and is found by integrating $p(T)$ from $T(D_{obs})$ to infinity. The Bayesian \textit{p}-value is the classical \textit{p}-value averaged over the posterior distribution of the model parameters. Small values of $p_B$ indicate that the model is very unlikely to generate the value $T(D_{obs})$. The reverse scenario, where the measured test statistic values $T_R$ and $T_{SSE}$ are very small compared to the distribution obtained from posterior predictive checking - and hence the associated $p$-values would be very large - would be an indication of overfitting, similar to obtaining anomalously small values in a $\chi^2$ test. In both scenarios, this is an indication that other models are required.

\subsection{Method summary}
\label{method_summ}

In summary, the analysis method presented here consists of the following steps.

\begin{itemize}
\item Acquire flare data and obtain the Fourier power spectral density (PSD). This PSD is henceforth referred to as the 'data' $D$.
\item Select candidate models $S$ with which to fit the data $D$.
\item Assign prior probabilities to the parameters of each model $S$.
\item For each model, perform MCMC simulations to find the posterior probability distribution and hence the best-fit to the data $D$.
\item Perform a model comparison using the $BIC$.
\item For the favoured model, perform posterior predictive checking to test the appropriateness of this model
\item Determine whether the chosen test statistics $T_{R}$ and $T_{SSE}$ are extreme.
\end{itemize}

Hence using this method we will have determined which of our tested models is a more appropriate description of that data, and additionally we will have quantified - using posterior predictive checking - how appropriate the preferred model is. In practical terms - and in relation to QPP - this will tell us whether the power spectrum of the events studies may be adequately described by a pure frequency-dependent noise model, or whether excess power corresponding to an oscillation may be present.

\section{Results}
\label{results}

\subsection{The flare of 2011 February 15}
\label{2011_feb_15}

Here we apply the method described above to the GOES-class X2.2 flare of 2011 February 15, which originated from NOAA active region (AR) 11158. This flare was previously analysed by \citet{2012ApJ...749L..16D}, who concentrated on the correlation between short-term variations in the signal at various wavelengths using data from a variety of instruments. An explicit significance test of the observed variations in the signal was not performed. This flare has also been studied by various other authors, including \citet{2011ApJ...738..167S, 2012ApJ...744..166T, 2014ApJ...783...98K, 2014ApJ...782L..31W}.

\begin{figure*}[ht]
\begin{center}
\includegraphics[width=19cm]{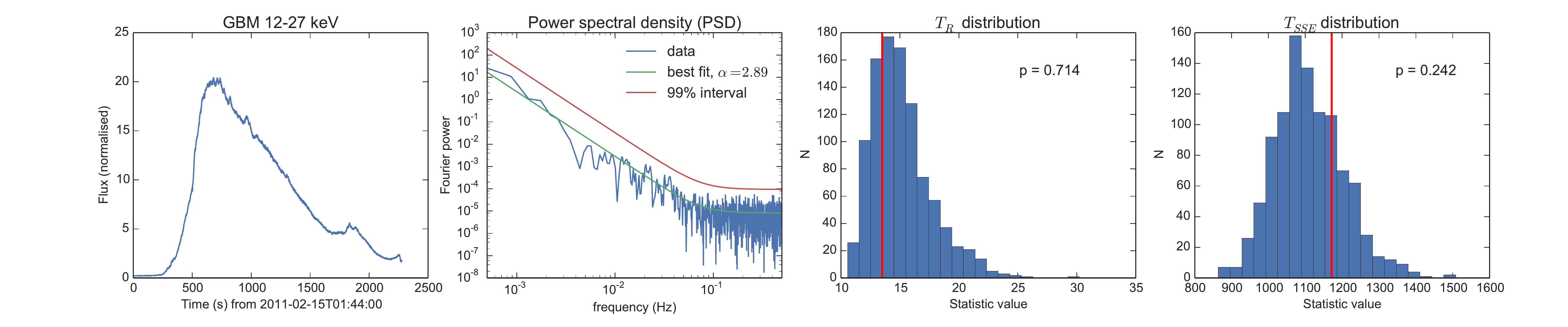}
\caption{Time series, power spectra and test statistic distributions $T_R$ and $T_{SSE}$ for Fermi/GBM observations of the 2011 February 15 solar flare. Each subfigure a) - d) shows the following: a) Original lightcurve in the 12-27 keV energy band, b) Fourier power spectral density of this signal after treatment with a window function, c) distribution of the test statistic $T_R$, d) distribution of the test statistic $T_{SSE}$. In panels c) and d) the observed value of the test statistic is denoted by the red vertical line.}
\label{example_analysis_fig}
\end{center}
\end{figure*}

For this event, the most sunward facing detectors of GBM show evidence of pulse pile-up effects due to high count rates. Therefore, we utilize detector NaI-2 for this analysis, which was pointed at a greater angle from the Sun, resulting in manageable counts. We apply the analysis method to the Fourier power spectra obtained from three X-ray energy ranges observed by GBM at 1 s cadence (12 - 27 keV, 27 - 50 keV, 50 - 100 keV), and to the Al and Zr filter data from LYRA, also at 1s cadence. Figure \ref{example_analysis_fig} shows an example of the method applied to the 12 - 27 keV data. The left panel shows the original signal, while the second panel shows the Fourier power spectrum. In this example, the simple model $S_A$ was strongly preferred over $S_B$, with $\Delta BIC$ = -14.4. The remaining two panels illustrate the $T_R$ and $T_{SSE}$ distributions obtained by posterior predictive checking as described in Section \ref{method_step3}, where the actual observed values of these statistics are denoted by the red lines. From this it can be seen that the observed values of $T_R$ and $T_{SSE}$ are not extreme within the context of the power law model. 

As Table \ref{table1} shows, the model comparison results for each X-ray band observed by GBM are similar, resulting in a negative value of $\Delta BIC$, with values ranging from -14.4 at 12 - 27 keV to -5.4 in the 50 - 100 keV range. Similarly, for both LYRA filters the value of $\Delta BIC$ is $< -10$. Hence the simple power law model $S_A$ is preferred in all cases over the more complex model $S_B$.

\begin{figure}
\begin{center}
%\vspace{-0.1cm}
%\includegraphics[width=8.3cm]{paper_fig_event_g415_n2_6.pdf}
\vspace{-0.1cm}
\includegraphics[width=8.3cm]{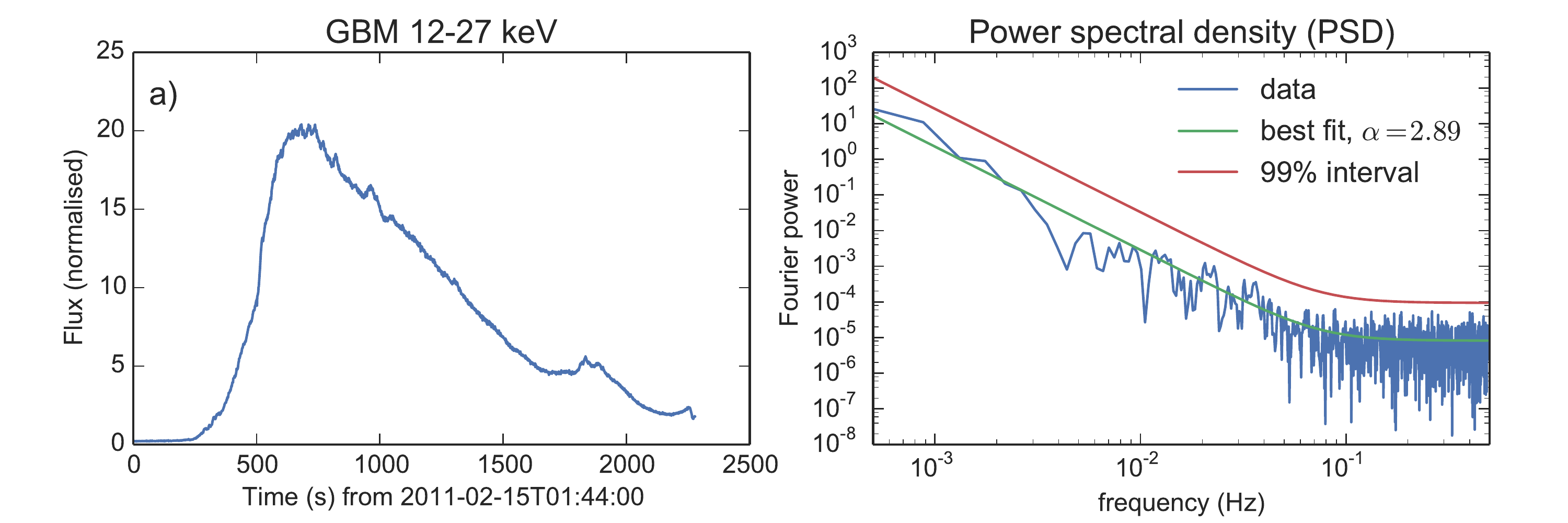}
\vspace{-0.1cm}
\includegraphics[width=8.3cm]{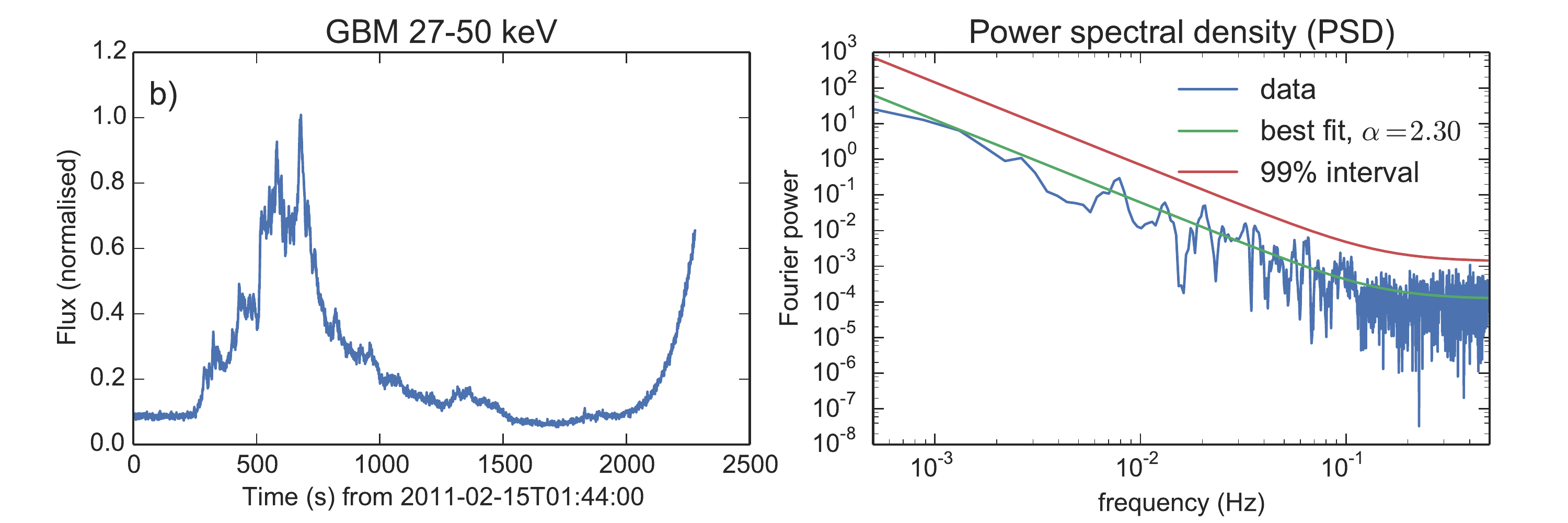}
\vspace{-0.1cm}
\includegraphics[width=8.3cm]{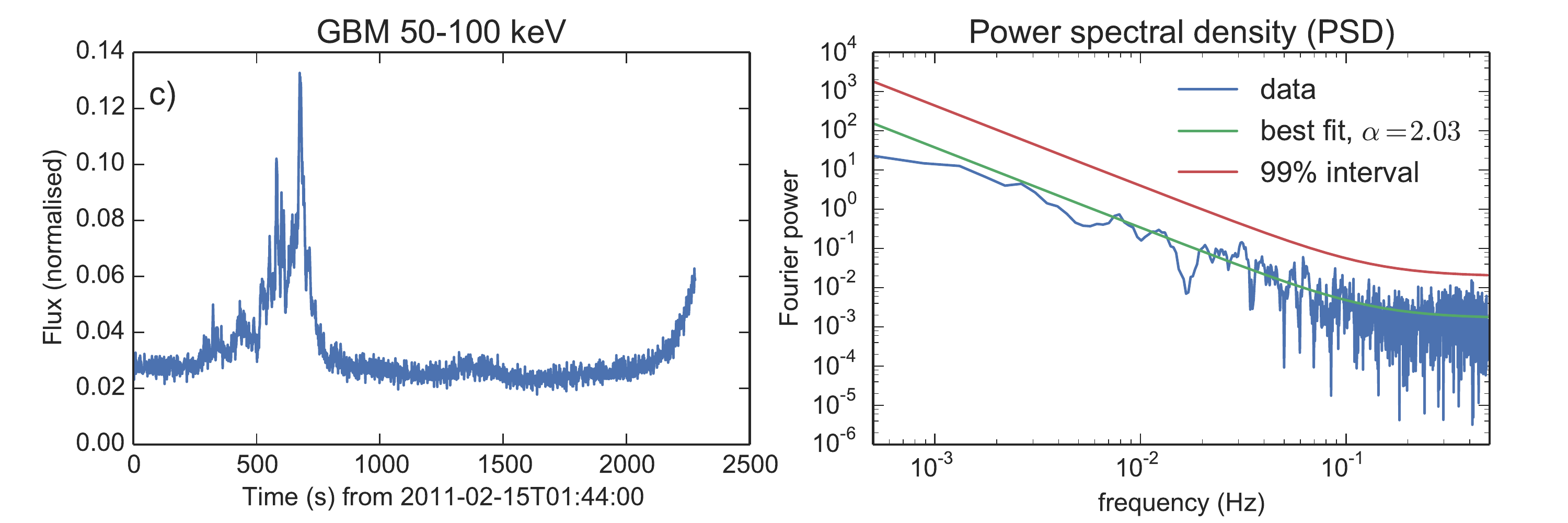}
\vspace{-0.1cm}
\includegraphics[width=8.3cm]{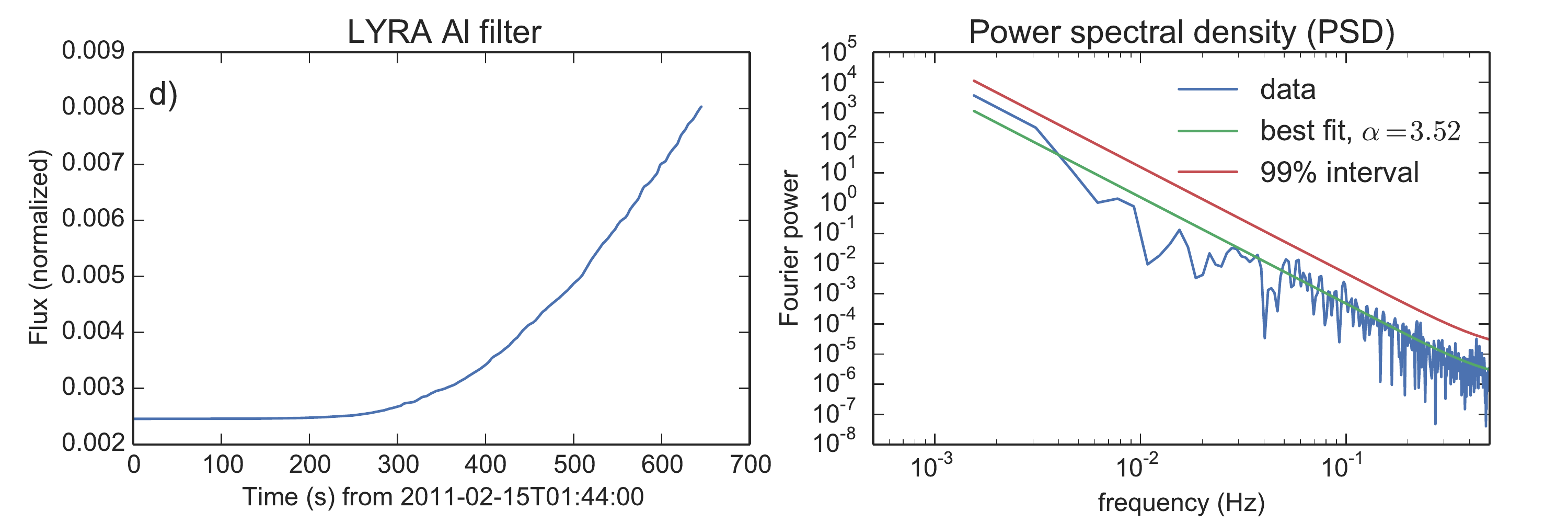}
\vspace{-0.1cm}
\includegraphics[width=8.3cm]{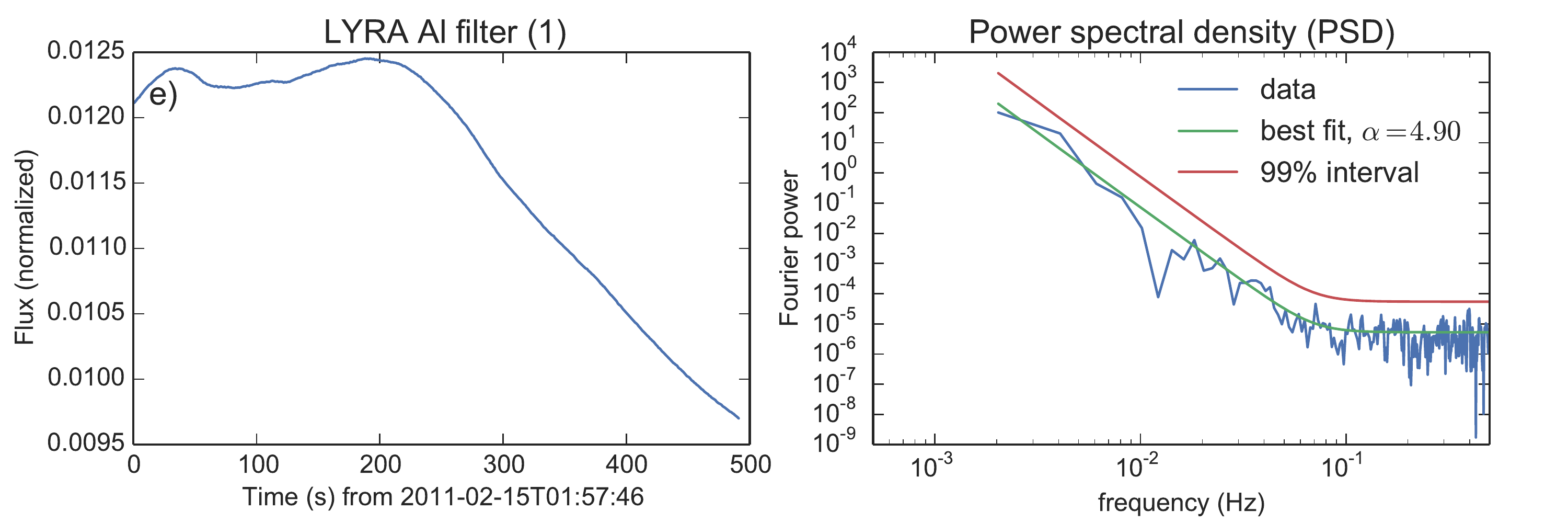}
\vspace{-0.1cm}
\includegraphics[width=8.3cm]{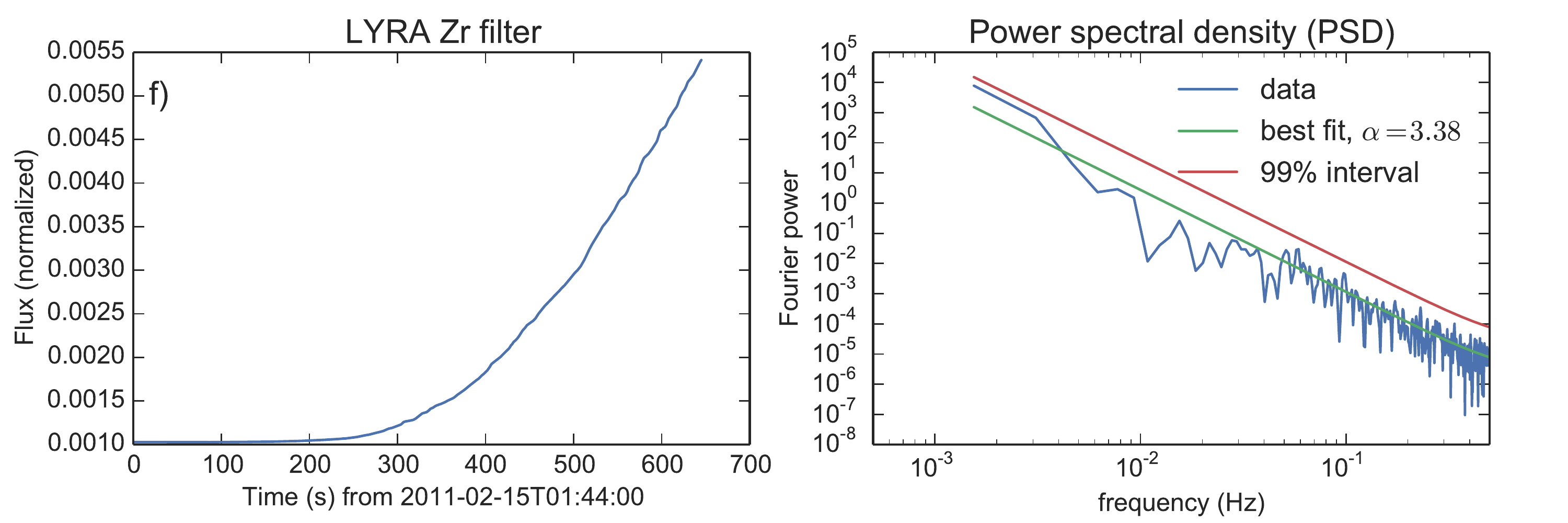}
\vspace{-0.1cm}
\includegraphics[width=8.3cm]{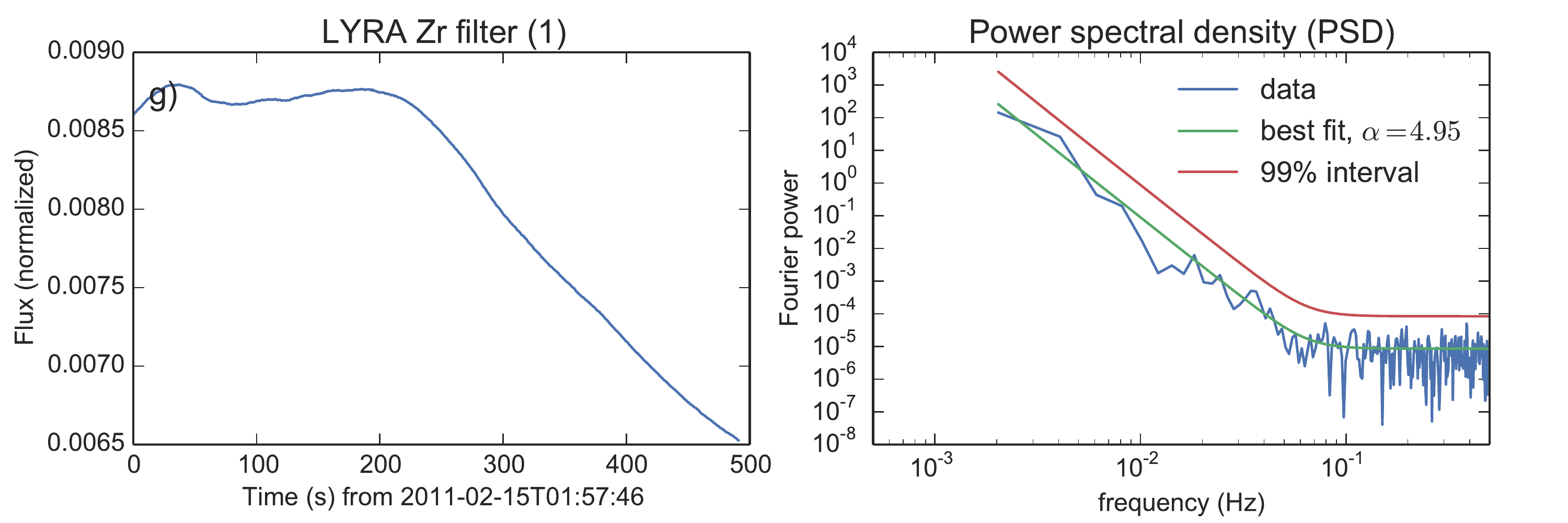}
\caption{Results of the analysis method applied at multiple wavelengths to the 2011 February 15 flare. \textit{Left column}: time series of the 2011 February 15 solar flare in three GBM X-ray bands (12-27 keV, 27-50 keV, 50-100 keV) and the LYRA Al and Zr channels. The LYRA data are split into two sub-series each due to the occurence of a LAR during this flare. \textit{Right column}: The Fourier power spectra of each time series (blue) and the associated best-fit model (green). Also shown is an estimate of the 99\% confidence level (red line) obtained by finding the value of $T_R$ at each frequency that would be consistent with a \textit{p}-value of 0.01. }
\label{2011_feb_15_fig}
\end{center}
\end{figure}

Figure \ref{2011_feb_15_fig} shows the lightcurves and best fits of model $S_A$ to the Fourier power spectra each of the investigated energy bands. For brevity the $T_R$ and $T_{SSE}$ distributions are not shown, but for reference each best-fit power law is accompanied by a 99\% significance line, which is obtained by finding the value of $T_R$ at each $f$ where the \textit{p}-value $p_{TR} < 0.01$. It can therefore be seen that the variations in the Fourier power spectra are not extreme in the context of the power-law model expectations. The \textit{p}-values for $T_{R}$ and $T_{SSE}$ are listed in Table \ref{table1}.

\begin{table*}
\caption{Summary of model comparison results and statistics in selected energy ranges for four flare events.}
\begin{center}
\scalebox{0.9}{
\begin{tabular}{c|ccccccccc}
\tableline
\tableline

Event & Fig. ref. & Obs. instrument & Obs. band & Obs. Interval (UT) & $\Delta$BIC & Preferred Model & -$\alpha$ & \textit{p}-value ($T_R$) & \textit{p}-value ($T_{SSE}$) \\
\tableline
\multirow{7}{*}{2011 Feb 15}  & 3b & GBM & 12 - 27 keV & 01:44 - 02:22 & -14.4 & $S_A$ & -2.89 $\pm$ 0.07 & 0.714 & 0.242  \\
					  & 3c & GBM & 27 - 50 keV & 01:44 - 02:22 & -10.7 & $S_A$ & -2.30 $\pm$ 0.06 & 0.261 & 0.077  \\
					  & 3d & GBM & 50 - 100 keV & 01:44 - 02:22 & -5.4 & $S_A$ & -2.03 $\pm$ 0.07 & 0.124 & 0.082 \\
					  & 3e &LYRA & Al (17-80 nm) & 01:44 - 01:55  & -14.3 & $S_A$ & -3.52 $\pm$ 0.12 & 0.177 & 0.021 \\
					  & 3f & LYRA & Al (17-80 nm) & 01:58 - 02:06 & -14.2 & $S_A$ & -4.90 $\pm$ 0.11 & 0.688 & 0.663 \\
					  & 3g & LYRA & Zr (6 - 20 nm) & 01:44 - 01:55 & -13.8 & $S_A$ & -3.38 $\pm$ 0.13 & 0.649 & 0.368  \\
					  & 3h & LYRA & Zr (6 - 20 nm) & 01:58 - 02:06 & -18.8 & $S_A$ & -4.95 $\pm$ 0.13 & 0.345 & 0.186  \\
\hline
\multirow{7}{*}{2011 Jun 7} 	  & 4b & GBM & 12 - 27 keV &06:16 - 06:58 & -18.2 & $S_A$ & -2.90 $\pm$ 0.07 & 0.990 & 0.688  \\
					  & 4c & GBM & 27 - 50 keV &06:16 - 06:58 & -9.6 & $S_A$ & -3.12 $\pm$ 0.07 & 0.928 & 0.932  \\
					  & 4d & GBM & 50 - 100 keV & 06:16 - 06:58 & -4.2 & $S_A$ & -2.81 $\pm$ 0.08 & 0.844 & 0.871 \\
					  & 4e & LYRA & Al (17-80 nm) &06:16 - 06:27 & -19.7 & $S_A$ & -3.48 $\pm$ 0.1 & 0.035 & 0.046 \\
					  & 4f & LYRA & Al (17-80 nm) &06:30 - 06:50 & -3.7 & $S_A$ & -2.96 $\pm$ 0.11 & 0.110 & 0.742 \\
					  & 4g & LYRA & Zr (6 - 20 nm)& 06:16 - 06:27 & -9.7 & $S_A$ & -3.46 $\pm$ 0.09 & 0.565 & 0.068 \\
					  & 4h & LYRA & Zr (6 - 20 nm)& 06:30 - 06:50 & -2.3 & $S_A$ & - 2.80 $\pm$ 0.10 & 0.007 & 0.014 \\
\hline
\multirow{4}{*}{1998 May 8}	  & 5 & NoRH & 17 GHz & 01:54 - 02:02 & 80.4 & $S_B$ & -3.06 $\pm$ 0.12 & 0.824 & 0.798  \\
					  & 5 & Yohkoh/HXT & 13-23 keV & 01:56 - 02:00 & 16.1 & $S_B$ & -3.70 $\pm$ 0.66 & 0.146 & 0.222 \\
					  & 5 & Yohkoh/HXT & 23-33 keV & 01:56 - 02:00 & -3.4 & $S_A$ & -2.27 $\pm$ 0.17 & 0.441 & 0.171 \\
					  & 5 & Yohkoh/HXT & 33-53 keV & 01:56 - 02:00 & -19.0 & $S_A$ & -1.86 $\pm$ 0.13 & 0.641 & 0.929 \\
\hline
2009 January 16		  & 6 & NMSU & U-band & 04:03 - 07:35 & -15.6 & $S_A$ & -3.55 $\pm$ 0.12 & 0.999 & 0.998  \\
'megaflare' & & & & &

\end{tabular}
}
\label{table1}
\end{center}
\end{table*}

\subsection{The flare of 2011 June 7}

This GOES-class M2.5 flare originated from AR 11226 and has been studied by various authors, including \citet{2012A&A...540L..10I, 2013ApJ...777...30I, 2013Sci...341..251R, 2013ApJ...776L..12G, 2014ApJ...782...87C}, primarily due to the spectacular prominence eruption associated with this event. However, this flare also exhibited strong pulsations in UV and X-ray wavelengths, which were investigated in the \citet{2013ApJ...777...30I} study. As with the event studied in Section \ref{2011_feb_15}, a statistical significance test was not performed on these pulsations, and their exact nature was left for debate in favour of timing studies and correlations with other flare parameters. 

\begin{figure}
\begin{center}
%\vspace{-0.1cm}
%\includegraphics[width=8.3cm]{paper_fig_event_g415_22.pdf}
\vspace{-0.1cm}
\includegraphics[width=8.3cm]{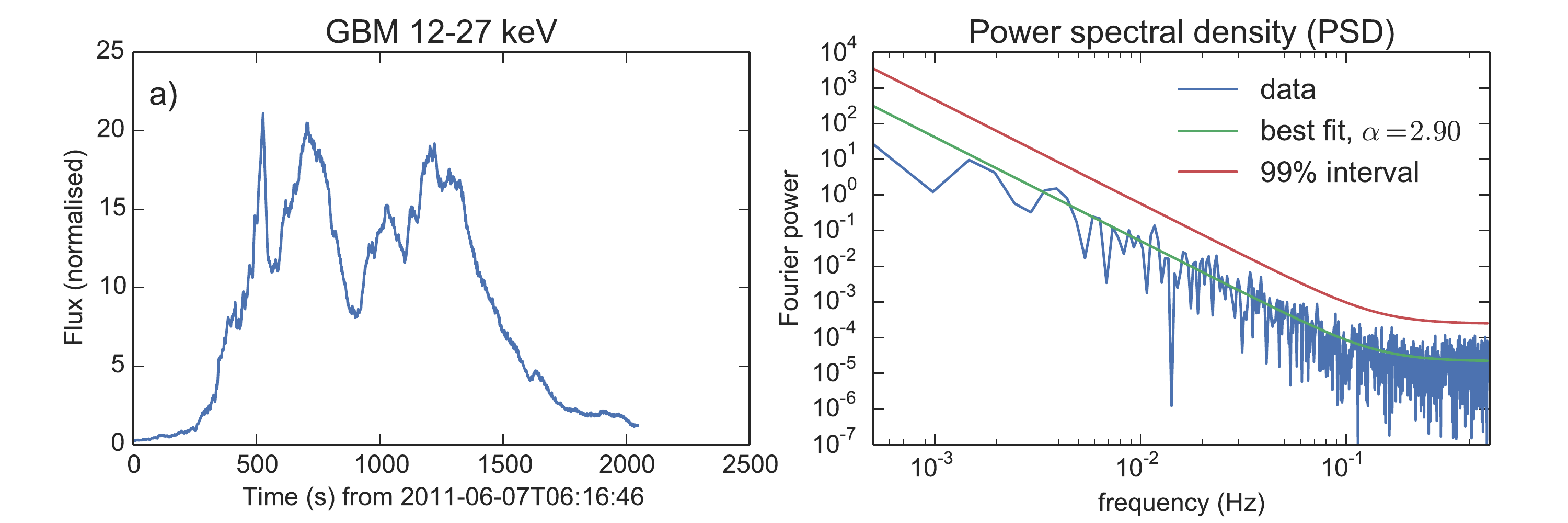}
\vspace{-0.1cm}
\includegraphics[width=8.3cm]{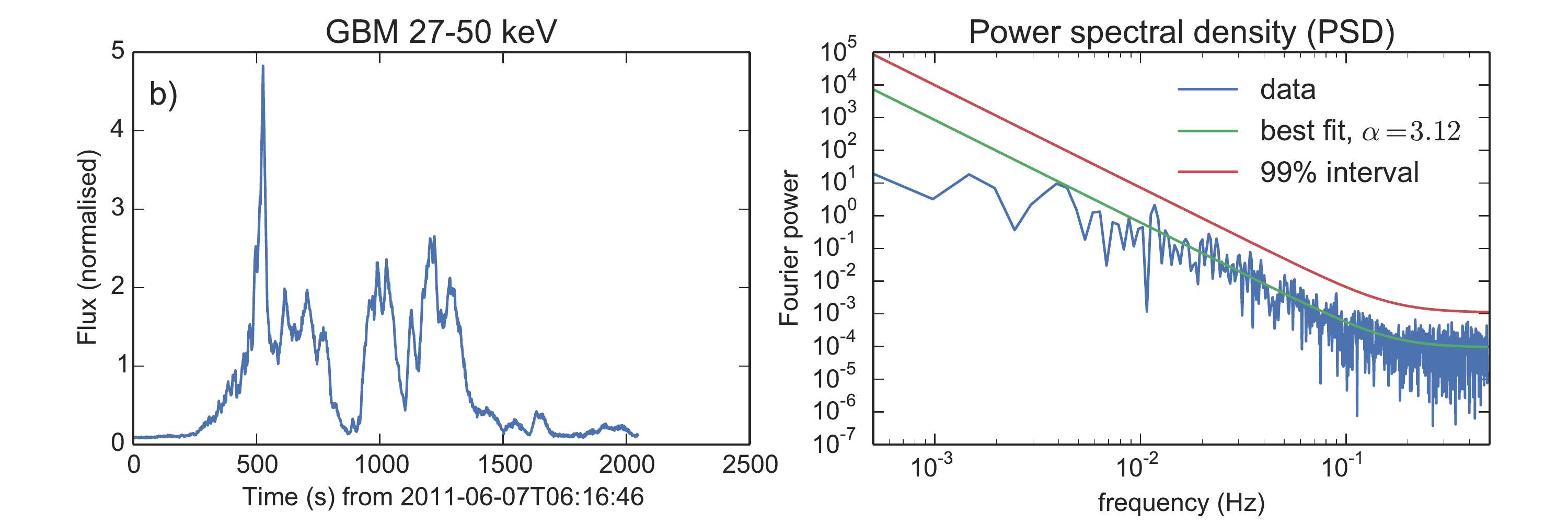}
\vspace{-0.1cm}
\includegraphics[width=8.3cm]{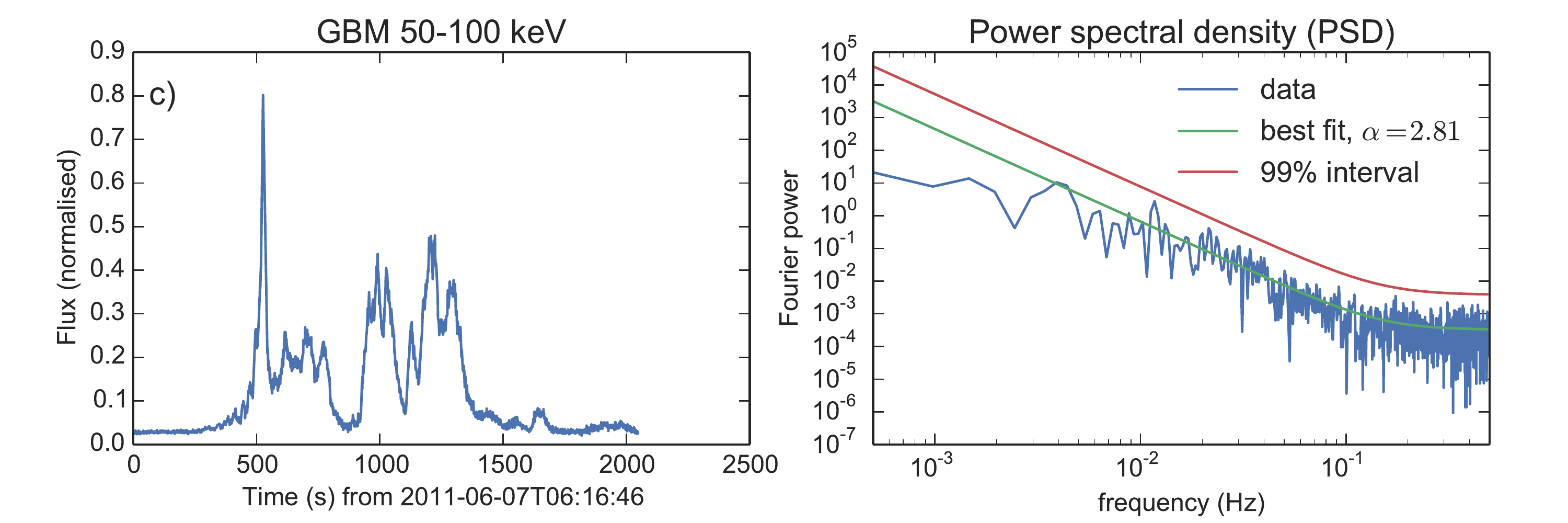}
\vspace{-0.1cm}
\includegraphics[width=8.3cm]{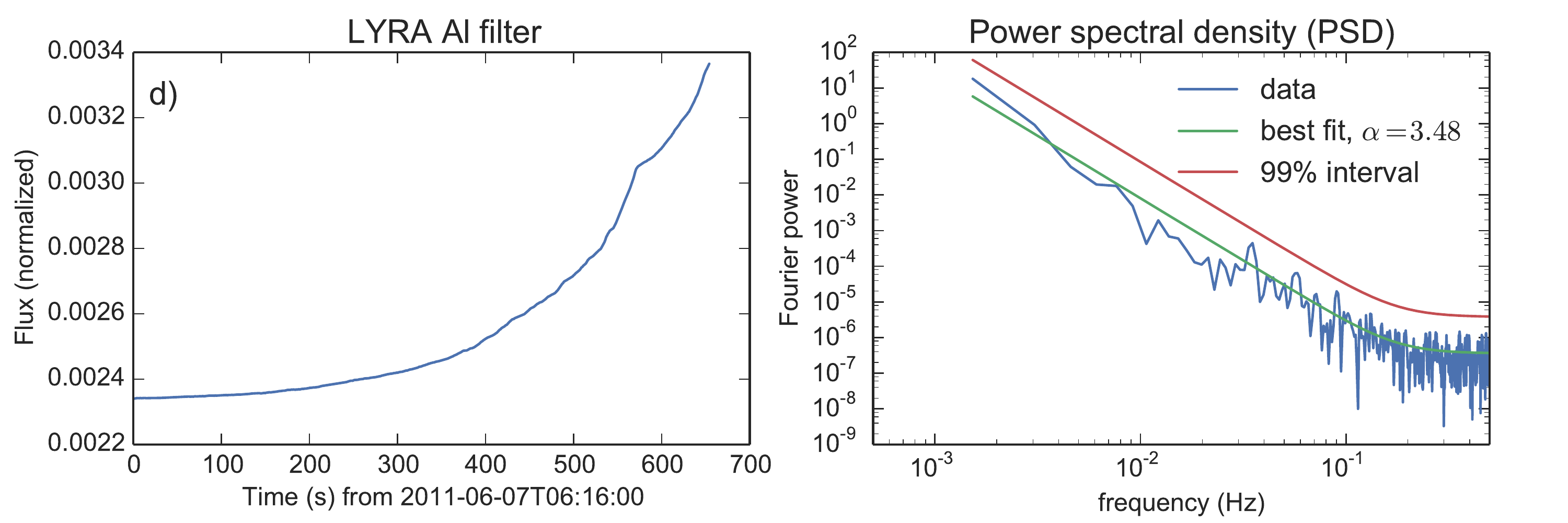}
\vspace{-0.1cm}
\includegraphics[width=8.3cm]{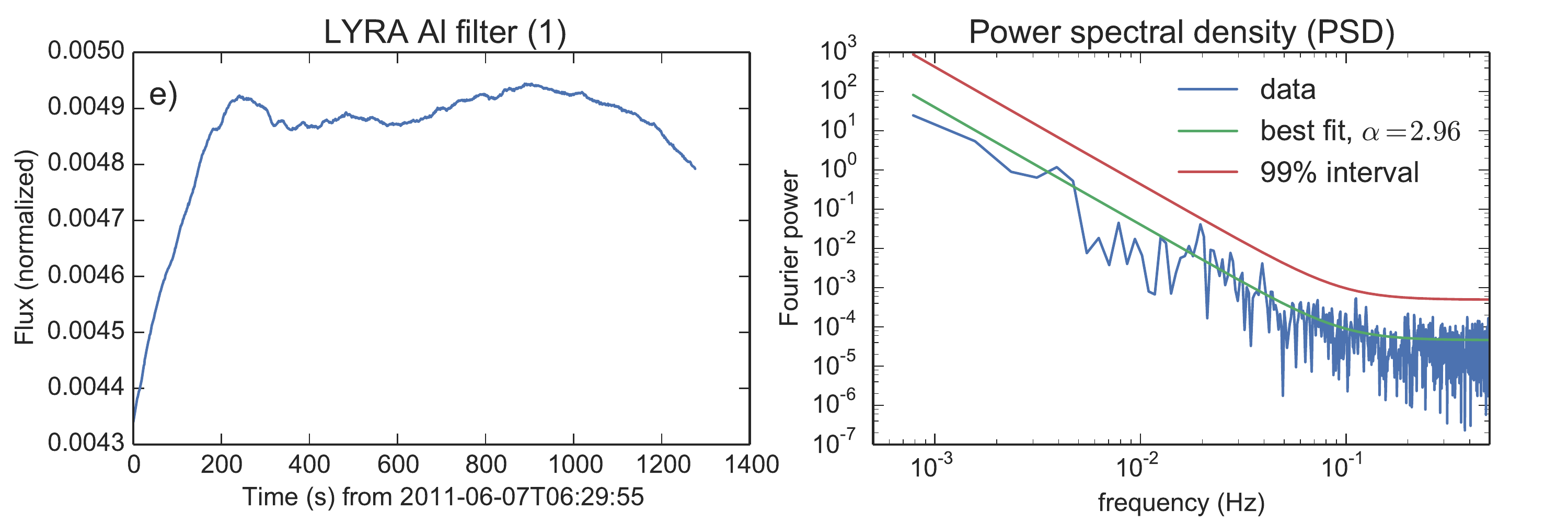}
\vspace{-0.1cm}
\includegraphics[width=8.3cm]{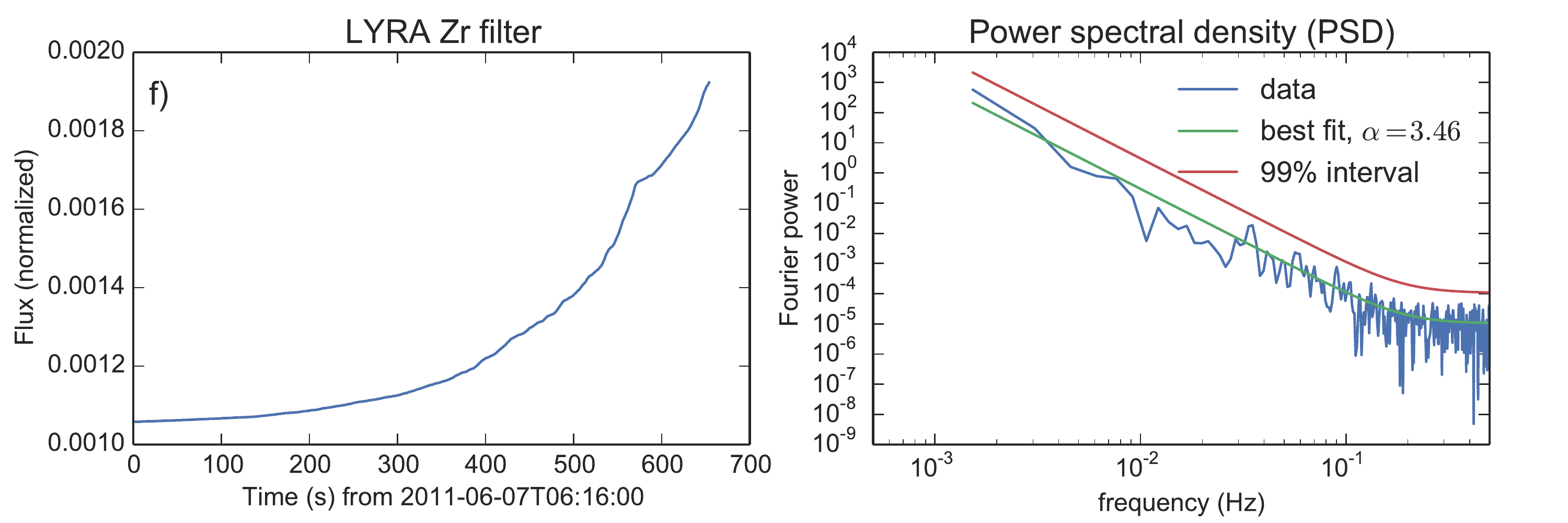}
\vspace{-0.1cm}
\includegraphics[width=8.3cm]{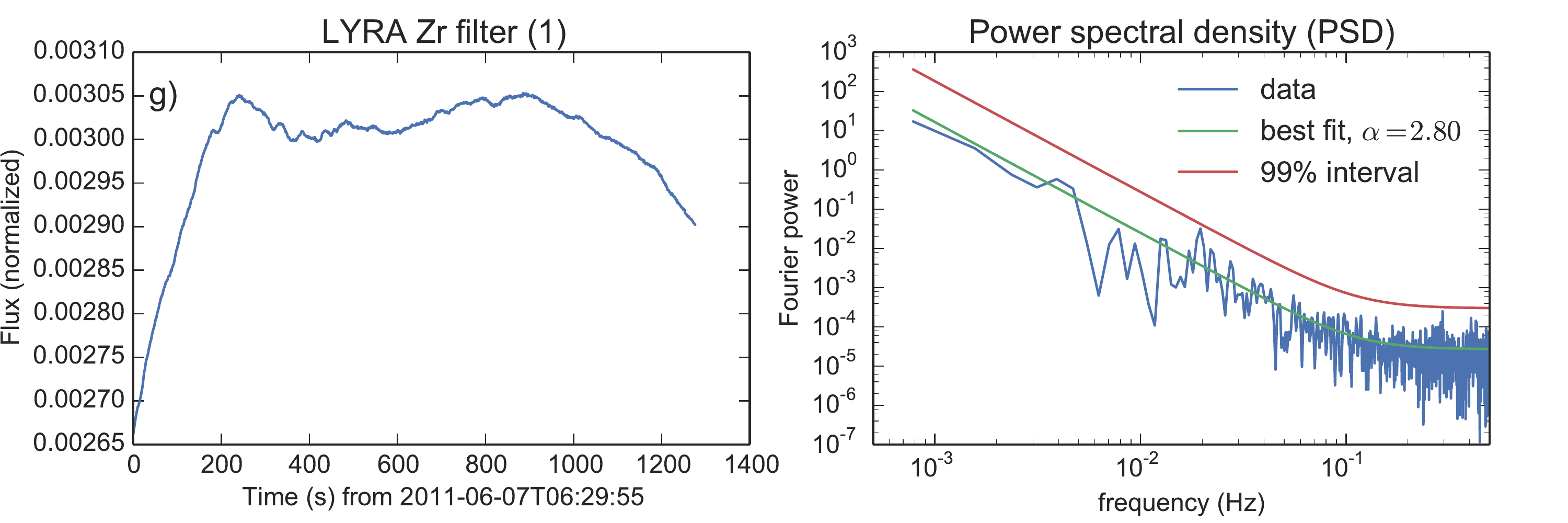}
\caption{Results of the analysis method applied at multiple wavelengths to the 2011 June 7 flare. \textit{Left column}: time series of the 2011 June 7 solar flare in three GBM X-ray bands (12-27 keV, 27-50 keV, 50-100 keV) and the LYRA Al and Zr channels. The LYRA data are split into two sub-series each due to the occurence of a LAR during this flare. \textit{Right column}: The Fourier power spectra of each time series (blue) and the associated best-fit model (green). Also shown is an estimate of the 99\% confidence level (red line) obtained by finding the value of $T_R$ at each frequency that would be consistent with a \textit{p}-value of 0.01.}
\label{2011_jun_7_fig}
\end{center}
\end{figure}

Figure \ref{2011_jun_7_fig} shows the result of considering power-law Fourier power spectra by applying the analysis method described in Section \ref{method}. For each of the six wavelengths studied, model $S_B$, which includes an additional spike that would be consistent with a narrowband oscillation, is never preferred over the pure power-law model $S_A$ (see Table \ref{table1} for the associated $\Delta BIC$ values), although in the late phase of the Al and Zr channel data there is little difference between the two models. Additionally, the test statistics $T_R$ and $T_{SSE}$ describing the goodness of fit for $S_A$, listed in Table \ref{table1}, are generally not extreme, indicating an overall good fit of the model $S_A$ to the data $D$ in most cases. An exception to this is the 12-27 keV data, where $p_{TR}$ is 0.01, suggesting that although $S_A$ is preferred over $S_B$, it is not a complete description of the observed data. Overall however, despite the visually striking pulses in the flare lightcurves, the observed power spectral densities at both X-ray and EUV wavelengths are consistent with and well-described by a power law model in the Fourier power spectrum, without the need for explicit oscillations in the model.

\subsection{The flare of 1998 May 8}

The GOES-class M3.1 flare of 1998 May 8 originated from AR 8210 and began at approximately 01:49 UT in soft X-rays. It was also observed in radio and hard X-rays by the Nobeyama Radioheliograph and the Yohkoh satellite respectively, and is a pronounced example of a QPP event. Analysis of this data in two previous studies \citep{2008A&A...487.1147I, 2004AstL...30..480S} suggested the presence of a statistically significant oscillation with P $\approx$ 16s, which was interpreted by \citet{2008A&A...487.1147I} as the manifestation of a magnetoacoustic sausage mode. However, these studies did not account for the power-law-like nature of the signal Fourier power spectrum, and hence may have overestimated the statistical significance of their results.

\begin{figure*}
\begin{center}
\includegraphics[width=4.2cm]{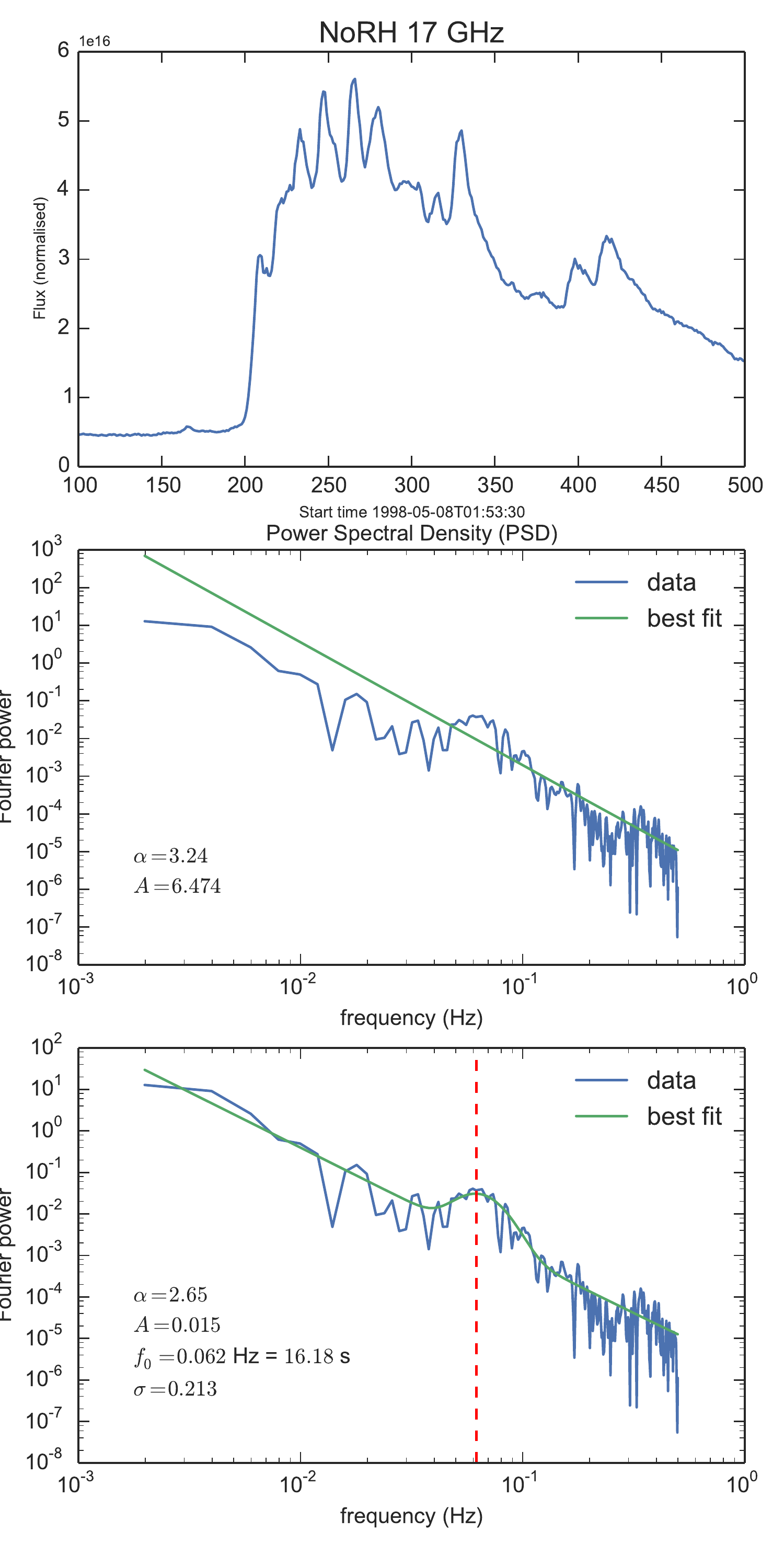}
\includegraphics[width=4.2cm]{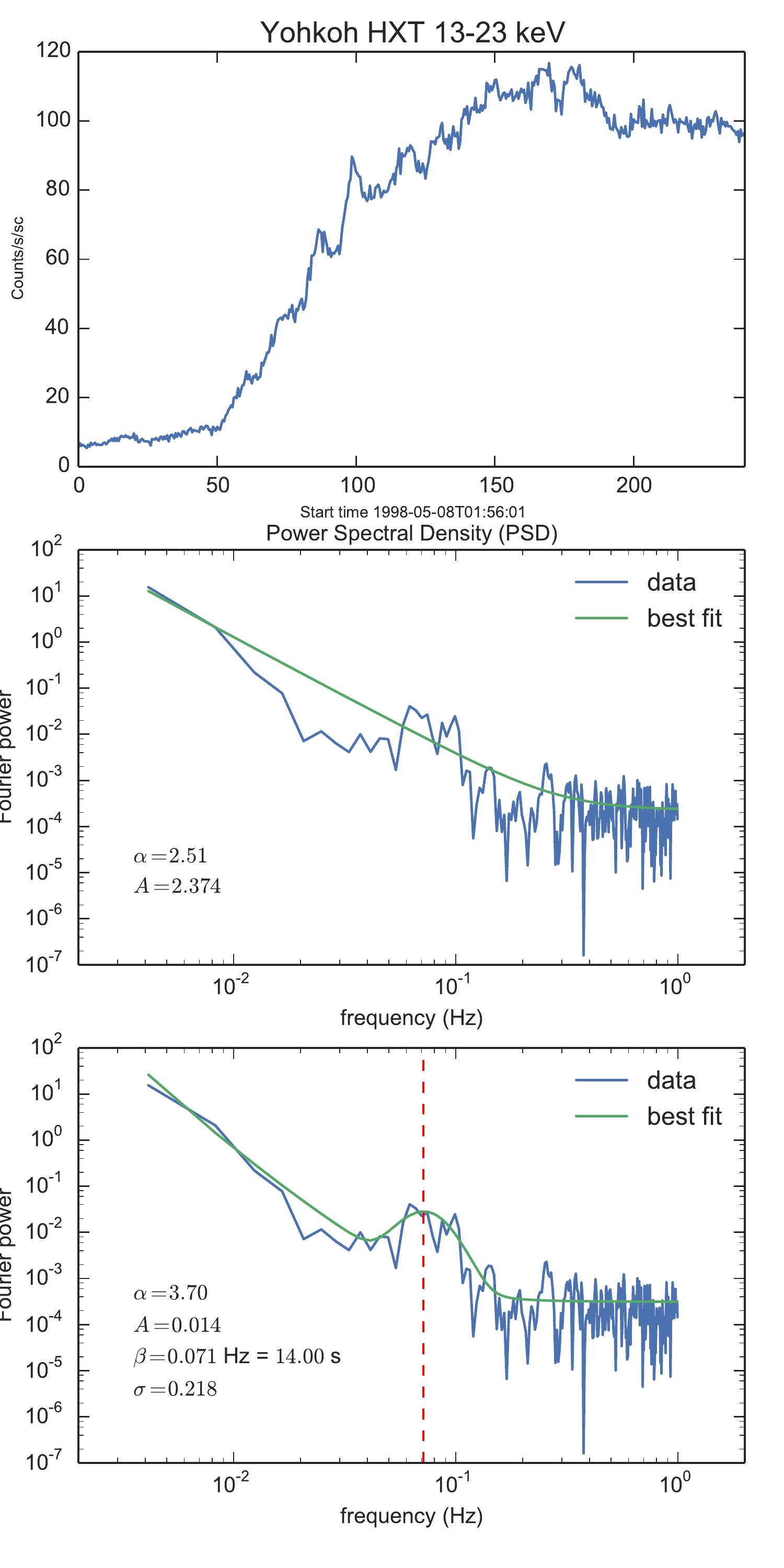}
\includegraphics[width=4.2cm]{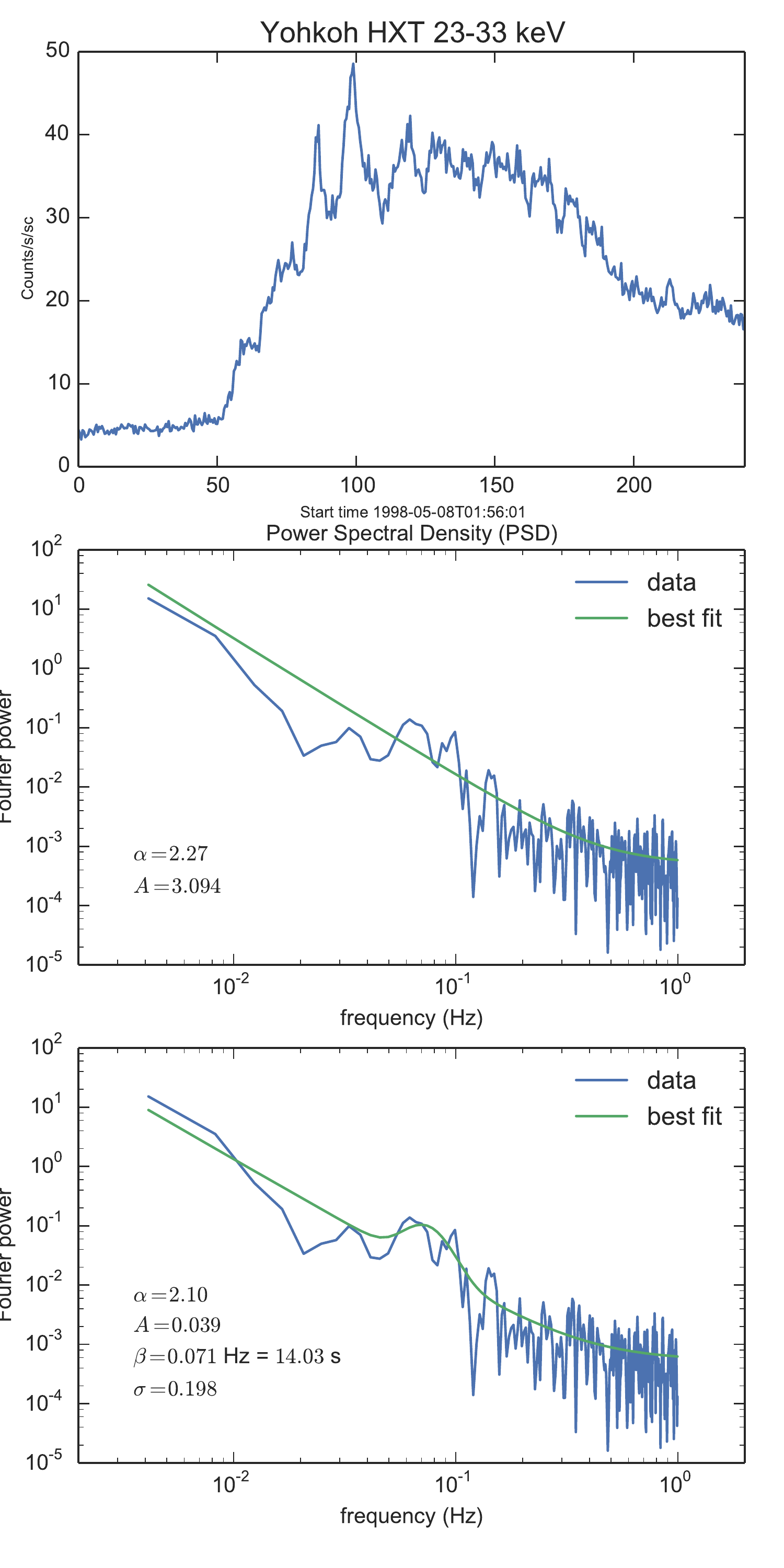}
\includegraphics[width=4.2cm]{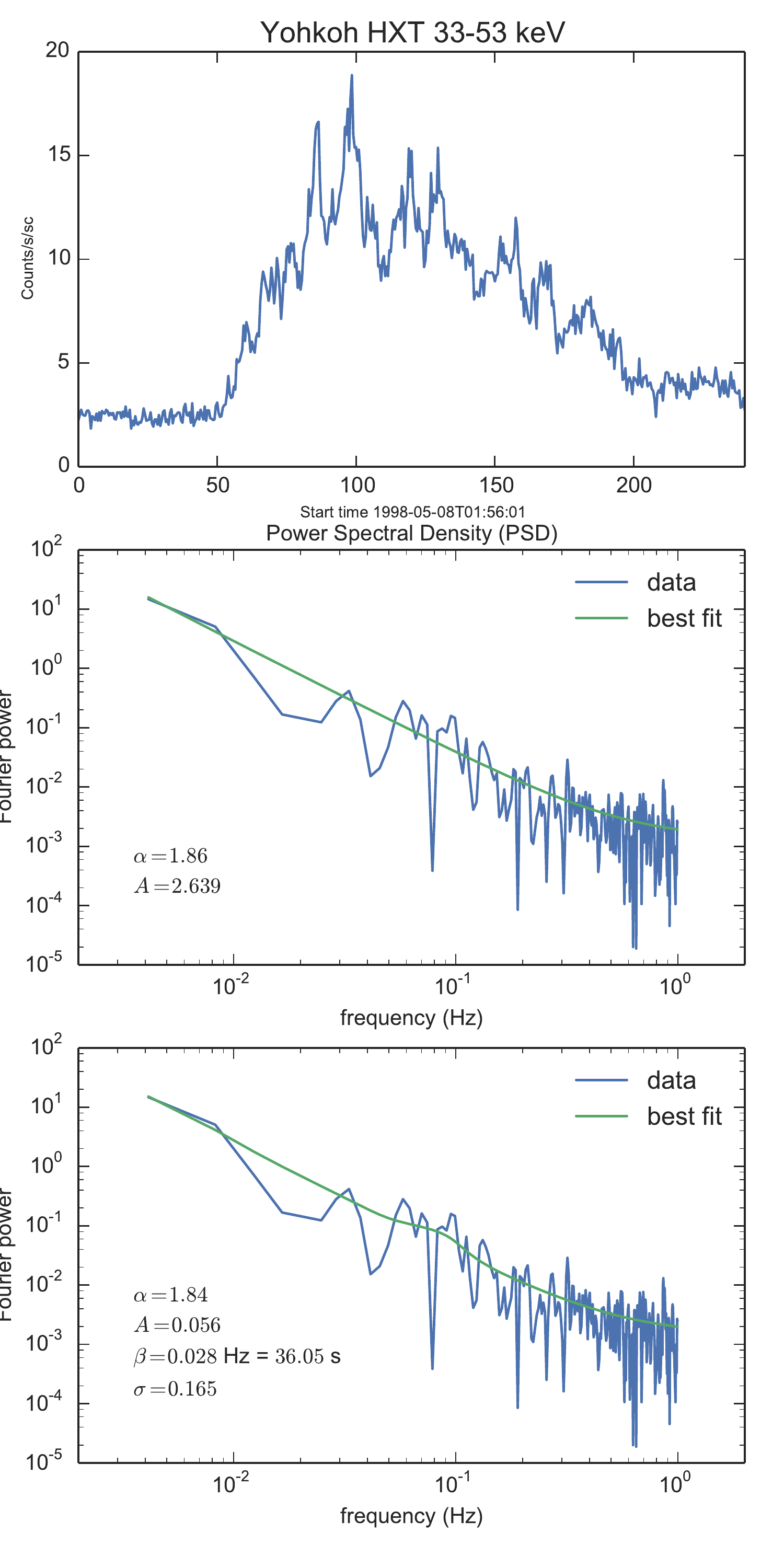}
\caption{Results of analysis method applied to the 1998 May 8 solar flare observational data from NoRH and Yohkoh. Columns from left to right: NoRH 17GHz radio data, Yohkoh L-channel (13-23 keV) data, Yohkoh M1-channel (23-33 keV) data, Yohkoh M2-channel (33-53 keV) data. For each column, the panel arrangement is as follows - top panel: the observed radio flux or X-ray counts associated with the flare. Center panel: the best fit of model $S_A$ (green) to the Fourier power spectrum (blue). Bottom panel: the best fit of model $S_B$ (green) to the Fourier power spectrum (blue). The $\Delta BIC$ values associated with these model comparisons are $\Delta BIC_{17GHz} \approx$84 for the NoRH 17GHz data, $\Delta BIC_{L} \approx$16 for the Yohkoh 13-23 keV signal, $\Delta BIC_{M1} \approx$ -3.4 for the Yohkoh 23-33 keV signal, and $\Delta BIC_{M2} \approx$ -19 for the Yohkoh 33-53 keV signal. Hence, model $S_B$ is strongly preferred for both the NoRH 17 GHz and Yohkoh 13-23 keV data, indicating an oscillation.}
\label{norh_event}
\end{center}
\end{figure*}

Revisiting this event, the method described in Section \ref{method_summ} is applied to both 1 s cadence Nobeyama Radioheliograph data at 17 GHz, and to 0.5 s cadence Yohkoh X-ray data. The results of the model comparison are shown in Figure \ref{norh_event}. We find that, for the 17 GHz emission, $\Delta BIC \approx$ 80, strongly favouring model $S_B$ which includes a Gaussian bump in addition to a power law. The best-fit location for this bump is at $f_0$ = 0.062 $\pm$ 0.003 Hz, equivalent to $P$ =  16.2 $\pm$ 0.8s. Although the position of the Gaussian bump is determined very accurately, in this context the width of the Gaussian is also of interest, as it encapsulates information about the range, or variation, of the observed `quasi'-period. Hence, when the Gaussian width is included in the period uncertainty we obtain $P$ = $16.2^{+3.8}_{-3.1}$ s. This is consistent with the results of \citet{2008A&A...487.1147I, 2004AstL...30..480S}, who both estimated that $P \approx$ 16s. 

In X-rays, analysis of the Yohkoh data in the 13-23 keV channel yields $\Delta BIC \approx$ 16, also indicating a strong preference for model $S_B$. The best-fit location of the frequency bump with a 1-$\sigma$ uncertainty is $f_0$ = 0.071 $\pm$ 0.004 Hz, corresponding to $P$ = 14.0 $\pm$ 0.75 s. As before, when we utilize the Gaussian width parameter to estimate the bounds of the quasi-period, we obtain $P$ = $14.0^{+3.4}_{-2.8}$ s, consistent with the observations at radio wavelengths. In the higher energy X-ray channels, the preference for model $S_B$ is not replicated - for the 23-33 keV channel, model $S_A$ is marginally preferred with $\Delta BIC \approx$ -3.4, while for the 33-53 keV channel model $S_A$ is strongly preferred with $\Delta BIC \approx$ -19. As Figure \ref{norh_event} shows however, the fitting of model $S_B$ to the 23-33 keV channel, where the simple power law is marginally preferred, yields the same best-fit period of $P \approx$ 14s as the 13-23 keV channel. This suggests that an oscillatory signature may indeed be present in the 23-33 keV data, but at insufficient strength to be unambiguously detected.

Examination of the test statistic values for this event (see Table \ref{table1}) reveals that the measured $T_R$ and $T_{SSE}$ values for the 17 GHz data are smaller than the mean of the distribution obtained via posterior predictive checking (see Section \ref{method_step3}), leading to \textit{p}-values of 0.842 and 0.798 respectively. For the 13-23 keV data we find the reverse effect, but in neither case are the test statistic values extreme. However, although a good fit was obtained here, it should be noted that our choice of a Gaussian function is a purely empirical one; as pointed out by \citet{2010MNRAS.402..307V}, the true statistics of the Fourier power spectrum in the region of the narrowband peak are unclear. Hence, although model $S_B$ is strongly preferred for both the 17 GHz and 13-23 keV data, indicating the presence of an oscillation with $P \approx$ 14 - 16s, the Gaussian function is likely not a statistically complete choice for modelling this oscillation.

The critical finding is that, for this event, a simple power-law Fourier power spectrum is not sufficient to explain the flare data at radio or soft X-ray wavelengths. Instead, there is strong evidence at both radio and X-ray energies for the presence of a narrow-band oscillation. This is an extremely important result, which confirms the existence of oscillation-like signatures in solar flare signals across multiple energies and emission regimes. This observation is consistent with classical ideas of QPP, such as bursty reconnection processes or MHD wave modes.

\subsection{The stellar megaflare of 2009 January 16}

As a final example, we consider the stellar megaflare of 2009 January 16, previously studied by \citet{2010ApJ...714L..98K} and \citet{2013ApJ...773..156A}. This flare originated from the dM4.5e star YZ CMi, and exhibited several emission peaks in addition to an exponential-like decay. These variations were interpreted by \citet{2013ApJ...773..156A} as a decaying long-period oscillation with $P \approx$ 32 minutes, suggesting a longitudinal magnetoacoustic mode as a possible mechanism. Here, we revisit the sub-minute cadence near-UV data from the NMSU 1m telescope studied by \citet{2010ApJ...714L..98K, 2013ApJ...773..156A}, and re-evaluate this flare signature accounting for power-law effects in the Fourier power spectrum. 

\begin{figure}
\begin{center}
\includegraphics[width=8.0cm]{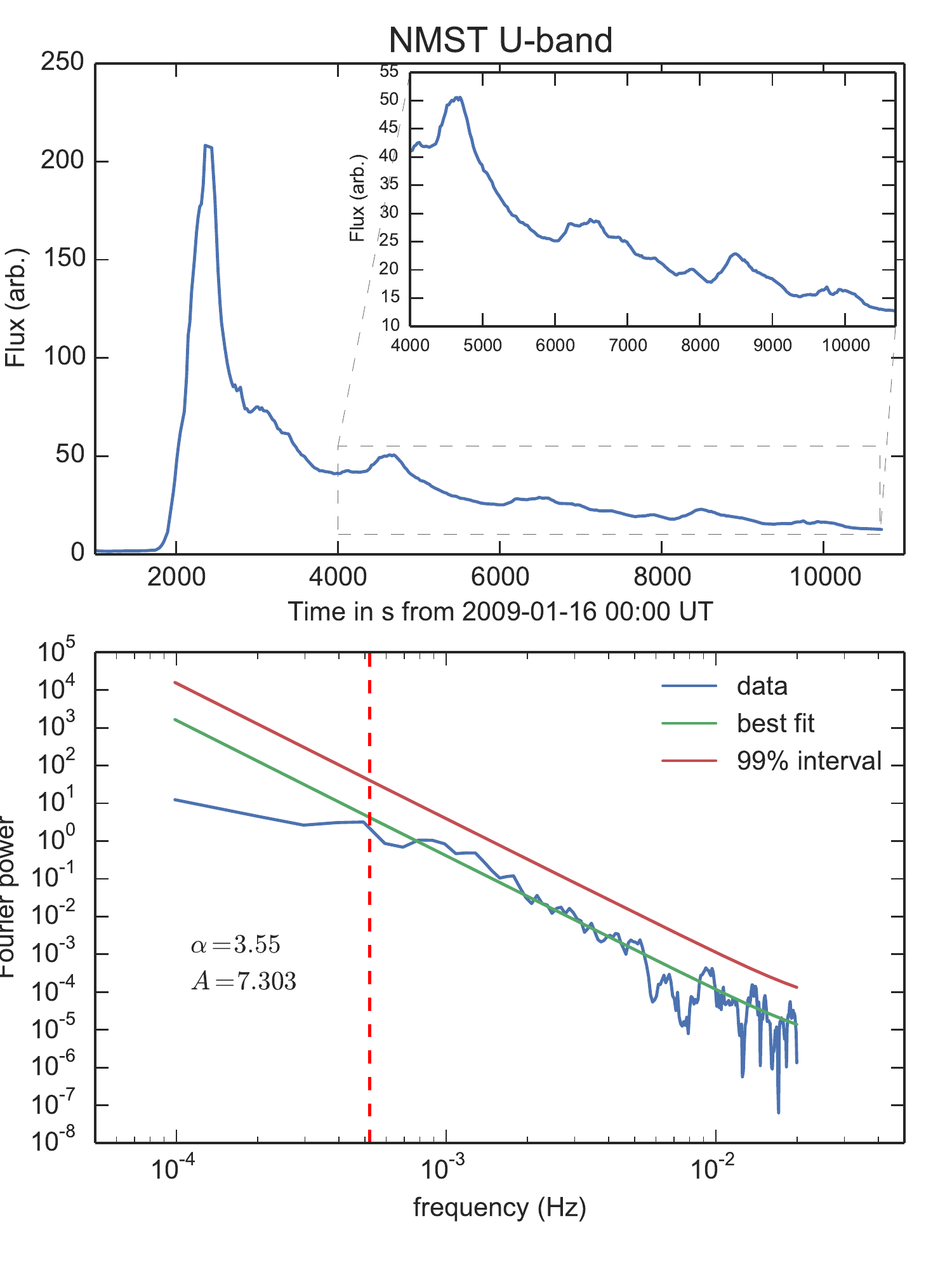}
\caption{Result of analysis method applied to the 2009 January 16 stellar `megaflare'. Top panel: The U-band lightcurve of the flare from dM4.5e star YZ CMi, as observed by the New Mexico State University 1m telescope. Bottom panel: The best fit of model $S_A$ (green) to the Fourier power spectrum (blue). For this event, $\Delta BIC$ $\approx -16$, indicating no evidence in favour of model $S_B$. The period of $P \approx$ 32 minutes suggested by \citet{2013ApJ...773..156A} is indicated by the vertical dashed line for reference. }
\label{megaflare}
\end{center}
\end{figure}

Figure \ref{megaflare} illustrates the U-band lightcurve and the best-fit Fourier power spectrum for this event. Here, model $S_A$ is preferred over $S_B$ by a substantial margin, with $\Delta BIC \approx$ -16. Hence, despite the intriguing variations in the observed time profile, the emission from this stellar flare is better explained in terms of the power law model than the model containing a narrowband oscillation. The observed $T_R$ and $T_{SSE}$ values are rather low however compared to their obtained posterior predictive distributions. This indicates that another untested model may better reproduce the statistics of the data than either $S_A$ or $S_B$.

\section{Discussion and interpretation}
\label{interpretation}

Having studied 19 time series from 4 flares, we have found strong evidence for the existence of an explicit oscillation only in the 17 GHz radio data and the 13-23 keV X-ray data from the 1998 May 8 flare. For the remaining events, the simple power law model $S_A$ was preferred as a description of the data over the model $S_B$ including an extra peak. However, in some cases the models were considered almost equally likely.  For example, the Al and Zr channel during the latter part of this flare (Figure \ref{2011_jun_7_fig}) showed $\Delta BIC \approx$ -3.7 and $\Delta BIC \approx$ -2.3 respectively, only marginally favouring $S_A$ over $S_B$. Hence, for these signals, we do not rule out the presence of an additional component - we can only say that we do not have strong evidence in favour of one. 

It is also important to emphasize that the adequate description of flares containing substantial temporal variations in terms of a power-law-like Fourier power spectrum does not mean that those variations are not real or do not exist. These variations are observed by multiple instruments and are of solar origin. Hence, studies of the timing of these variations, their variations across wavelengths and energy regimes \citep[e.g.][]{2012ApJ...749L..16D}, and their correlation with other measurable flare parameters \citep[][]{2013ApJ...777...30I} remain crucial towards our understanding of flares. The results of Section \ref{results} indicate that, when the frequency-dependent properties of flares are properly considered, an explicit oscillatory signal is not required in order to explain these observations. Instead, we must consider these variations as an intrinsic property of the flare system.

\begin{figure}
\begin{center}
\includegraphics[width=8.5cm]{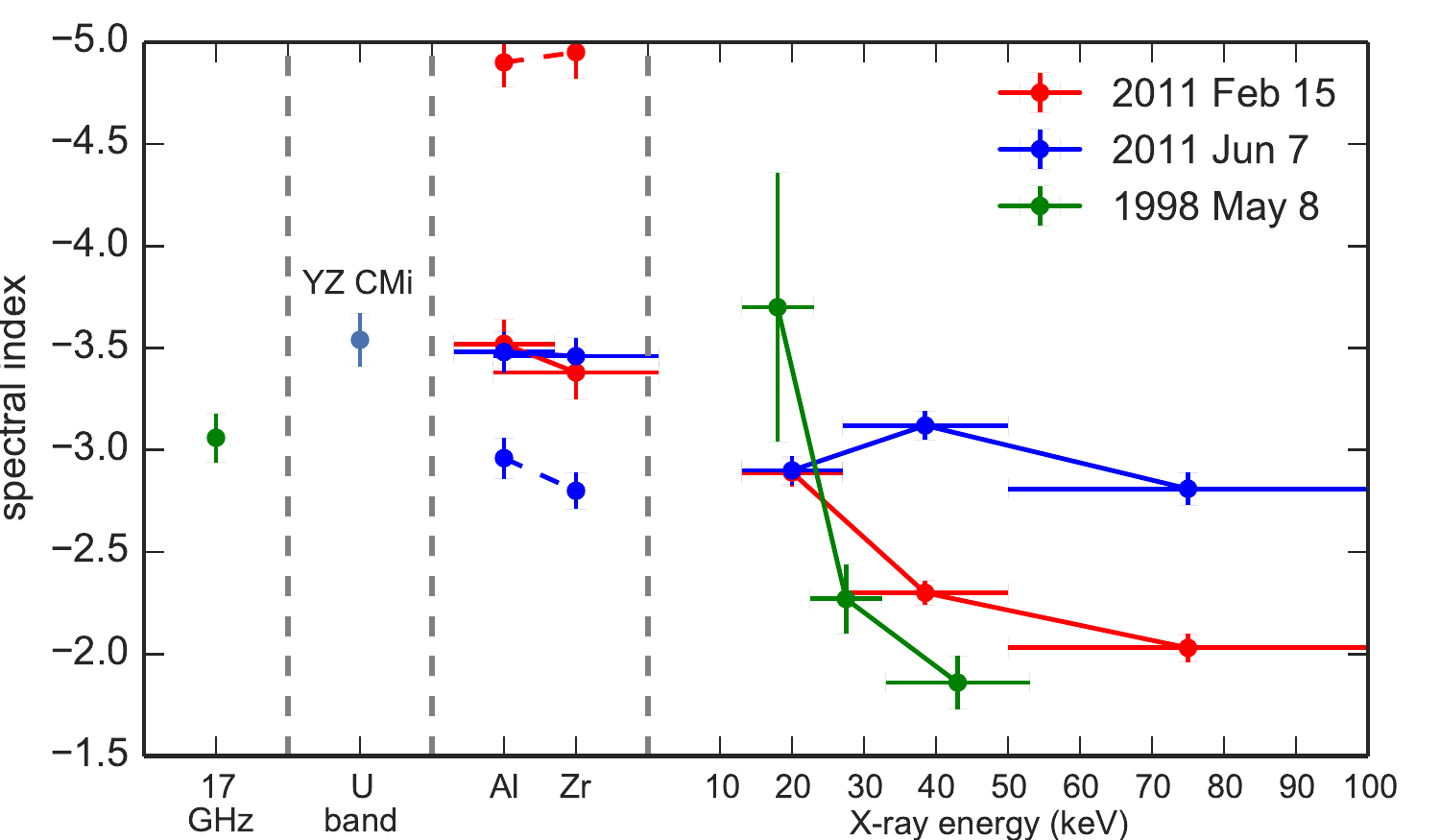}
\caption{Best-fit spectral indices $\alpha$ to the Fourier power spectra for the 2011 February 15 (red), 2011 June 7 (blue), and 1998 May 8 (green) events as a function of wavelength. The measured spectral index for the 2009 January 16 megaflare is indicated by the cyan data point. Note that this plot encompasses different emission and energy regimes, which are separated by vertical lines for clarity. The dashed red and blue lines highlight the LYRA power law indices for the later segments of the two flares respectively. }
\label{spectral_indices}
\end{center}
\end{figure}

One possible mechanism that would give rise to a power law structure in solar flares was suggested by \citet{2011soca.book.....A}. In this model, a flare is considered as a superposition of many exponentially decaying energy deposition events. If the number of events is $N(E) \propto E^{-\alpha_{E}}$, and the energy in each event is $E(T) \propto T^{1+\gamma}$, where $T$ is exponential decay time, then the resulting Fourier spectrum should have a power law index

\begin{equation}
p = -(2 - \alpha_E)(1+\gamma)
\label{aschwanden_eqn}
\end{equation}

Hence the observed power law index can in principle provide information about the energy release process in a flare.

\citet{2007ApJ...662..691M} also investigated the nature of bursty solar X-ray emission in the context of self-organised criticality and fractal behaviour. In that work, the H{\"o}lder exponent was determined - using RHESSI data - as a function of X-ray energy for the X4.8-class flare of 2002 July 23. The H{\"o}lder exponent is related to the Fourier power spectral index and is a measure of the persistence - i.e. how later observed values in a signal depend on earlier values - of the physical system. In a persistent walk regime, a generally increasing signal means that the next observed value in time is also likely to show an increase (`smooth behaviour'), whereas in an anti-persistent regime the opposite occurs, and a decrease is likely (`bursty behaviour'). In general, it was found that, for the 2002 July 23 flare, the persistence of the signal decreased as a function of energy; low energy bands (e.g. 6-12 keV) were consistent with a persistent-walk regime, while at higher energies (e.g. 50-100 keV) anti-persistent behaviour became more pronounced. This was also reflected in the measured Fourier power law exponents; the power law index $\alpha$ ranged from $\approx$ -2.6 at 3-6 keV down to $\approx$ -1.8 at 300 - 800 keV, indicating different signal characteristics.

Figure \ref{spectral_indices} shows the distribution of measured $\alpha$ values for the 2011 February 15, 2011 June 7, and 1998 May 8 events as a function of energy. The 2009 January 16 flare measurement is also shown. For 2011 February 15, the results of \citet{2007ApJ...662..691M} are reproduced; the measured spectral index in hard X-rays monotonically decreases as a function of energy from $\approx$ -2.9 in the 12-27 keV range, down to $\approx$ -2.0 at 50-100 keV. Similarly, the three datapoints available from Yohkoh/HXT data on 1998 May 8 show a monotonic decrease versus energy, although it should be noted that the spectral index is very poorly constrained in the 13-23 keV channel. The 2011 June 7 flare however displays inconsistent results: the measured exponents remain approximately constant at $\approx$ -3.0 as a function of energy in X-rays (see Figure \ref{spectral_indices}). This suggests that flares may not universally follow the monotonic trend in spectral index and fractal dimension observed by \citet{2007ApJ...662..691M}; different flares may exhibit different emission characteristics as a function of energy. This is not unexpected given the earlier results of \citet{1998ApJ...505..941A} who, in a large study of events in $>$ 25 keV X-rays observed with CGRO/BATSE, found substantial variation of the power law index between flares (-1.5 $< \alpha <$ -3.2 for strong events).

In interpreting these results, we should be wary of potential biases that may affect the measurement of $\alpha$. One potential issue is a dependence of $\alpha$ on the total fluence of the data, however this possibility is mitigated by the treatment of the input signal described by Equation \ref{norm_eqn}. Additionally, the choice of GBM detector for the 2011 February 15 and 2011 June 7 events ensures that pile-up effects should not be an issue in the X-ray data. The other main factor in determining $\alpha$, the length of the power law component in the Fourier power spectrum, is captured by the uncertainties listed in Table \ref{table1}. In general we find that $\alpha$ is well and consistently constrained, with the notable exception of the 13-23 keV X-ray data from Yohkoh in the 1998 May 8 flare. In this case, the power law slope is very poorly constrained due to the relatively early transition to a white-noise regime (see Figure \ref{norh_event}) and the location of the Gaussian bump.

An additional feature of these results is that not only do the observed Fourier power law indices vary between flares \citep[e.g.][]{1998ApJ...505..941A}; the observed power law index of both LYRA channels changes substantially between the two sub-intervals of both the 2011 June 7 and 2011 February 15 flares, as highlighted in Figure \ref{spectral_indices}. This could be related to the different physical phases of solar flares, e.g. the early 'rise phase' that includes impulsive X-ray emission, and the long decay phase often observed at EUV wavelengths. For the 2011 February 15 event, the power law index is much steeper in both the Aluminium and Zirconium filter channels during the observed decay phase of the LYRA emission compared with the earlier rise phase. In the context of Equation \ref{aschwanden_eqn}, this may indicate a change in either the number distribution of heating events, or a change in the energy distribution of these events, which would be expected during a transition to the flare decay phase. For the 2011 June 7 event, the difference in observed Fourier power law index is much less pronounced (Figure \ref{spectral_indices}) in the LYRA channels. One difference is that in this case, the later sub-intervals shown in Figures \ref{2011_jun_7_fig}e, g are associated with a peak of observed EUV emission, rather than a strong decay phase, which is not observed due to a large-angle rotation of the LYRA spacecraft (see Section \ref{lars}).

\section{Conclusions}

It is clear that power laws in the Fourier power spectrum are an intrinsic property of both solar and stellar flare time series, and one that must be taken into account when searching for pulsations or oscillations in flares. As has been shown here, events which may be previously considered as containing signatures of oscillations can in fact be adequately described via a power-law model of the Fourier power spectrum. Hence, the prevalence of explicit oscillations in flares is likely to have been substantially overestimated in past literature, although few large-sample studies exist. However, the analysis of the 1998 May 8 flare shows that, for at least some events, there is strong evidence of oscillatory signals in the data over and above the background frequency-dependent spectrum, in both radio and X-ray emission regimes. In these cases, classical interpretations of QPP, such as regimes of bursty reconnection or signatures of magnetohydrodynamic wave modes, remain valid. Additionally, as discussed in Section \ref{interpretation}, the detection and measurement of frequency-dependent Fourier spectra may provide us with a diagnostic tool for understanding the fundamental processes of flare enery release. The measured power law index as a function of energy provides us with information regarding the energy release process in different energy regimes.

These results are relevant not just to flare studies, but throughout solar physics, and coronal seismology in particular, where many searches for waves and oscillations implicitly assume a Gaussian noise regime. For example, \citet{2014A&A...563A...8A} recently observed that, for a number of regions of interest, the Fourier power spectra of long-duration EUV data obtained by SOHO/EIT also obey a power law. Similarly, \citet{2014AAS...22432321I} reveal Fourier power-law properties in the time series of active regions obtained in EUV by SDO/AIA. Hence, power-law Fourier power spectra must be accounted for in many solar features over many timescales and in multiple emission regimes.

A future large-scale study of this kind is required in order to understand how the Fourier power law index may vary as a function of energy between flares, during the flares themselves, and how this information may be used to diagnose conditions at the energy release sites in flares. Such a study will also establishing the true prevalence of explicit oscillations in solar and stellar flares.

\begin{acknowledgements}
ARI and JI both gratefully acknowledge the support of the PROBA2 Guest Investigator Programme, which provided the opportunity and funds for both authors to collaborate with the PROBA2 team at the Royal Observatory of Belgium. The authors are also grateful to Dr Adam Kowalski for providing the U-band data from the NMSU 1m telescope. This work was carried out using the SunPy, SciPy, PyMC, and Matplotlib software packages. We also acknowledge the anonymous referee for useful comments which resulted in a number of improvements to this paper.
\end{acknowledgements}

\bibliographystyle{apj}
\bibliography{ingl0815}

\end{document}